\documentclass[%
 reprint,
nofootinbib,
 amsmath,amssymb,
 aps,
prb,
]{revtex4-2}
\usepackage{array}
\usepackage[usenames,dvipsnames]{xcolor}
\usepackage{graphicx}
\usepackage{dcolumn}
\usepackage{bm}
\usepackage[dvipsnames]{xcolor}
\usepackage[colorlinks=true,citecolor=Cerulean,linkcolor=RubineRed,urlcolor=Cerulean]{hyperref}
\usepackage{color}
\usepackage{mathtools}
\usepackage{amsmath}
\usepackage{braket}
\usepackage{mathrsfs}
\newcommand{\ep}[1]{^{(#1)}}

\newcommand{\av}[1]{\langle #1 \rangle}
\newcommand{\Av}[1]{\left \langle #1 \right \rangle}

\newcommand{\mc}[1]{\mathcal{#1}}
\newcommand{\mr}[1]{\mathrm{#1}}
\newcommand{\ts}[1]{\textsc{#1}}

\newcommand{\ha}[1]{\hat{#1}}

\newcommand{\pablo}[1]{\textcolor{black}{#1}}

\begin{document}

\graphicspath{{../images/}}

\title{Decomposing the Spectral Form Factor}

\author{Pablo Martinez-Azcona}
\thanks{These authors contributed equally to the work}

\affiliation{Department of Physics and Materials Science, University of Luxembourg, L-1511 Luxembourg, Luxembourg}

\author{Ruth Shir}
\thanks{These authors contributed equally to the work}

\affiliation{Department of Physics and Materials Science, University of Luxembourg, L-1511 Luxembourg, Luxembourg}

\author{Aur\'{e}lia Chenu}
 \email{aurelia.chenu@uni.lu}
\affiliation{Department of Physics and Materials Science, University of Luxembourg, L-1511 Luxembourg, Luxembourg}

\date{\today} 

\begin{abstract}
Correlations between the energies of a system's spectrum are one of the defining features of quantum chaos. They can be probed using the Spectral Form Factor (SFF). We investigate how each spectral distance contributes in building this two-point correlation function. Specifically, starting from the spectral distribution of $k$-th neighbor level spacing ($k$nLS), we provide analytical expressions for the $k$-th neighbor Spectral Form Factor ($k$nSFF). We do so for the three Gaussian Random Matrix ensembles and the `Poissonian' ensemble of uncorrelated energy levels. 

We study the properties of the $k$nSFF, namely its minimum value and the time at which this minimum is reached, as well as the energy spacing with the deepest $k$nSFF. This allows us to quantify the contribution of each individual $k$nLS to the SFF ramp, which is a characteristic feature of quantum chaos. In particular, we show how the onset of the ramp, characterized either by the \textit{dip} or the \textit{Thouless} time, 
shifts to shorter times as contributions from longer-range spectral distance are included. 
Interestingly, the even and odd neighbors contribute quite distinctively, the first being the most important to built the ramp. They respectively yield a resonance or antiresonance in the ramp. 

All of our analytical results are tested against numerical realizations of random matrices. We complete our analysis and show how the introduced tools help characterize the spectral properties of a physical many-body system by looking at the interacting XXZ Heisenberg model with local on-site disorder that allows transitioning between the chaotic and integrable regimes.


\end{abstract}

\maketitle


\section{Introduction}
Quantum many-body systems are studied in fields as diverse as condensed matter, statistical mechanics, quantum information, and high-energy physics. Although their spectra are usually too complicated to be described and studied analytically, certain spectral statistical properties are universally shared among different systems. These spectral statistical properties have become a probe of whether a quantum system is chaotic or integrable, motivated by the Bohigas-Giannoni-Schmit \cite{bohigas_characterization_1984} and the Berry-Tabor \cite{berry_level_1977} conjectures, respectively.
More specifically, the eigenvalues in a quantum chaotic system show level repulsion and thus are correlated. 
The distribution of nearest-neighbor Level Spacings (nnLS) in quantum chaotic systems follows closely the prediction from Random Matrix Theory (RMT), and, depending on certain basic symmetries, fall into one of the three universality classes of random matrices \cite{wigner_conference_1956, wigner_1951, Gaudin_1961, Wigner1967, berry_chaology_87, gutzwiller_periodic_71, mehta2004random, haake_quantum_1991}. Hence, the study of random matrices has become central to the study of quantum chaos. 
The existence of spectral correlations extends beyond nnLS and can be probed by other measures---including the spectral rigidity, the number variance, and the spectral form factor (SFF) \cite{haake_quantum_1991, wimberger_nonlinear_2014}--- which can be computed analytically using random matrix theory \cite{mehta2004random}. By contrast, integrable systems exhibit no such correlations---a property that can be attributed to the existence of an extensive number conserved charges in such systems \cite{caux2011remarks}. They can be described by the Poisson ensemble.

In this work we focus on the SFF. It is defined as the Fourier transform of the two-point correlation function and probes the spectrum of a given quantum system. 
 For chaotic systems, the time evolution of the SFF displays an initial system-dependent decay (the \textit{slope}), reaches a minimum in the \textit{dip}, starts growing back with a universal \textit{linear ramp}, and finally saturates to a constant value---the \textit{plateau}. The SFF contains information of all correlations between the energies, and we will show that all of them play a role in building the linear ramp.  This characteristic evolution is taken as a signature of chaos, since integrable systems do not exhibit a linear ramp---they can decay directly to the plateau or show periodic oscillations in time. Regardless of whether the system is chaotic or not, at finite temperatures the early time decay of the SFF is universally bounded \cite{martinez-azcona_analyticity_2022}. Also, the SFF can be used to set a universal bound on the quantum dynamics, which gives insight into thermalization and scrambling \cite{vikram2022exact}. 
Note that the SFF is not self-averaging \cite{prange_spectral_1997}, so its computation requires an average over an ensemble or a small time window. %

Using the $k$-th neighbor level spacing ($k$nLS) known in the literature \cite{abul-magd_wigner_1999}, we analytically find the contribution of each spectral spacing to the SFF, i.e. the $k$-th neighbor Spectral Form Factor ($k$nSFF). Specifically, we find that for RMT the $k$nSFF can be written as a Gaussian envelope and an oscillating function, part of the latter capturing the non-Gaussianity of the $k$nLS distributions, for small $k$, since at large $k$ the distributions are Gaussian \cite{brody1981random}. Conversely, for integrable systems described by the Poissonian ensemble, it is given by a Lorentzian envelope with a different oscillating function. 
We compute analytically the minimum value and corresponding time of each of the $k$nSFF, as well as the neighbor range $k^*$ which shows the deepest $k$nSFF. These quantities exhibit different behavior in the RMT and Poisson ensembles. Furthermore, we study how longer-range spectral correlations contribute to building the universal ramp of the SFF. To this end, we introduce the partial SFF, which has a cutoff in the neighbor range $K$. To see how this quantity progressively builds the ramp as $K$ increases, we study two time scales: the \textit{dip} time, defined as the time after which the relative maxima of the partial SFF grow, and the \textit{Thouless} time \cite{Edwards_1972, suntajs_quantum_2020}, defined as the time after which the partial SFF remains close to the universal connected SFF. 

We provide analytical results for the three Gaussian ensembles, namely the Orthogonal (GOE), Unitary (GUE), and Symplectic (GSE) ensembles, whose spectral statistics define the three universality classes of Hermitian random matrices.
For the spectral statistics of integrable systems, we provide results for the `Poissonian' ensemble, with uncorrelated random energies taken from a uniform distribution. Our analytical results are based on tractable approximations and compared to numerical realizations of RMT.  We illustrate our results in a physical system: the XXZ spin chain with disorder, which exhibits a transition from chaos to integrability as a function of the disorder's strength. This model is often used in the context of many-body localization  \cite{vznidarivc2008many, pal2010many, serbyn2016spectral, bertrand2016anomalous, sierant2019level, sierant2020model} and its SFF has been studied in \cite{ sierant2020model, suntajs_quantum_2020, schiulaz2019thouless, schiulaz_self-averaging_2020}.

The paper is structured as follows:  Section \ref{sec:model} briefly reviews some key results in random matrix ensembles that are relevant for our analysis. We introduce and explicitly compute the contribution of the $k$nLS to the Spectral Form Factor ($k$nSFF) in Sec.  \ref{sec:knSFF_results}, and study its properties in Sec. \ref{sec:properties_knSFF}. Sec. \ref{sec:building_SFF_fromknSFF} presents how each of the $k$nLS distributions participate to building the SFF, showing the progressive build-up of the correlation hole as the number of neighbors is increased. We find a surprising resonance when separating the even or odd neighbors' contributions. Section ~\ref{sec:XXZ} illustrates the introduced concepts in a physical model known to exhibit a transition from chaos to integrability, the XXZ spin chain with local on-site disorder. A discussion and conclusions in light of the current results are presented in Sect.~\ref{sec:discussion}. 

The Appendices provide further details on the numerical and analytical results and include: further details on the dip and Thouless times (cf. App. \ref{app:dip_time_SFF}), the standard results of the connected SFF (cf. App. \ref{App:SFF_connected}), details on the unfolding procedure (cf. App. \ref{App:unfolding}), a discussion on the lack of self-averaging of $k$nSFF's (cf. App. \ref{app:self-av}), a test of the approximations used in the analytical $k$nSFF's and how they affect the full SFF (cf. App. \ref{App:SFF_test}). In addition, we present a toy model with only nearest-neighbor correlations according to the Wigner distributions of level spacings, in which we observe that the SFF does not exhibit a linear ramp (cf. App. \ref{App:SFF_test}). Finally, we propose a dissipative protocol to measure correlation functions related to the $k$nSFF (cf. App. \ref{sec:protocol_knSFF}). 

\section{Random Matrix Ensembles \label{sec:model}}
We start by reviewing some key results for random matrices belonging to either the Gaussian random ensembles or the Poisson ensemble. 

First, we consider Gaussian random matrices. These can be described by the joint probability density of its $N$ eigenvalues \cite{mehta2004random}
\begin{eqnarray} \label{JPD_eigenvalues}
    \rho_\beta(E_1,\dots, E_N)= C \!\!\! \prod_{1\leq i <j\leq N} \!\! |E_i-E_j|^\beta \, e^{-A\,\sum_{i=1}^N E_i^2},\quad
\end{eqnarray}
where $\beta$ is the Dyson index distinguishing the different ensembles, taking the values $\beta=1,2,4$ for the GOE, GUE, GSE, respectively. The constant $C$ is not relevant in this work and $A$ sets the energy scale, which we keep free for now. A random matrix taken from a Gaussian ensemble is simply constructed by sampling each matrix element from Gaussian distributions, and enforcing the appropriate symmetry of the ensemble. 

To look at energy correlations, we start from the distribution of the $k$th neighbor level-spacing ($k$nLS), $s_n^{(k)}=E_{k+n}-E_n$. This is formally defined for a random matrix ensemble of $N\times N$ matrices as a generalization of the expression for the nnLS distribution in a $3\times 3$ random matrix  \cite{Atas_2013, kahn_statistical_1963}
\begin{align}\label{Pk_dist-definition}
    \mathcal{P}_\beta\ep{k}(s) &= \int_{-\infty}^\infty \!\! \mr dE_1 \int_{E_1}^\infty \!\! \mr dE_2 \, \dots  \int_{E_k}^\infty \!\! \mr dE_{N} \rho_\beta(E_1,\dots,E_{N})\nonumber\\
    & \times \frac{1}{N-k}\sum_{n=1}^{N-k} \delta\big[s-(E_{k+n}-E_n)\big].
\end{align}
One can take also a similar approach to Wigner's idea for the nnLS surmise and study the smallest matrix containing a $k$ spacing, i.e. setting $N=k+1$ \cite{rao_higher-order_2020}.
For $k=1$, the distribution of nearest-neighbor level spacings $\{s_i^{(1)}\}_{i=1}^{N-1}$ of the unfolded spectrum are well approximated by the Wigner surmise distribution 
    $\mathcal{P}\ep{1}_\beta(s) = C_{\beta,1} \, s^{\beta} \, e^{-A_{\beta,1} s^{2}}$, where  $C_{\beta,1}$ and $A_{\beta,1}$ are constant in $s$ and depend only on $\beta$. 
The probability distribution for any energy spacing $s\ep k$ has been the focus of various studies \cite{rao_higher-order_2020, engel_higher-order_1998, abul-magd_wigner_1999, tekur_higher-order_2018, sakhr_wigner_2006, forrester_random_2009}. 
The first generalization of Wigner surmise that we are aware of assumes a Brody-like ansatz, which essentially leaves the power-law parameter, $\alpha$ in $s^\alpha$, as a free parameter \cite{engel_higher-order_1998}. In Ref. \cite{abul-magd_wigner_1999}, the power law is found using a small $s$ expansion and the generalized Wigner surmise, Eq. \eqref{Pk_dist} below, is obtained assuming a Gaussian behavior at large $s$. This approach is also followed in \cite{sakhr_wigner_2006} in the context of 2D Poisson point processes. These references propose a generalization of Wigner surmise in the form
\begin{equation}\label{Pk_dist}
    \mathcal{P}_\beta\ep{k}(s) \approx C_{\alpha} \,s^{\alpha}\, e^{-A_{\alpha}s^2}.
\end{equation}
The parameter $\alpha$ depends on the spectral distance $k$ and the ensemble index $\beta$ through $\alpha=\frac{k(k+1)}{2}\beta+k-1$. 
 The values of $A_{\alpha}$  and $C_{\alpha}$ are 
 \begin{equation}\label{eq:AandC} 
A_{\alpha} = \left[\frac{\Gamma\left(\frac{\alpha}{2}+1\right)}{k\, \Gamma\left( \frac{\alpha+1}{2} \right)}\right]^2, \:\: 
     C_{\alpha} = \frac{2}{\Gamma\left(\frac{\alpha+1}{2} \right)} \left[\frac{\Gamma\left(\frac{\alpha}{2}+1\right)}{k\, \Gamma\left(\frac{\alpha+1}{2} \right)}\right]^{\alpha+1}\hspace{-1.5em}.
\end{equation}

Second, we consider random matrices taken from the Poissonian ensemble, where each matrix is diagonal with elements sampled from a uniform distribution.
We label this ensemble with $\beta=0$ for convenience. In this ensemble, the $k$nLS distribution reads\footnote{Note that in Engel \textit{et al.} \cite{engel_higher-order_1998}, $k_\textsc{emv}$ refers to the number of levels in between two levels, e.g. for the nnLS $k_\textsc{emv}=0$ instead of $k=1$ in our setting. Thus, the notations are consistent for $k_\textsc{emv} = k-1$.} \cite{engel_higher-order_1998} 
 \begin{eqnarray}\label{Pk_Pois}
    \mathcal{P}_{0}\ep{k}(s) &=& \frac{1}{(k-1)!}\, s^{k-1} e^{-s}~.
 \end{eqnarray}


\section{Contributions of $k$-th neighbor correlations to the SFF}\label{sec:knSFF_results}

The spectral form factor at infinite temperature is defined in terms of the system's energies $\{E_i\}_{i=1}^N$ as
\begin{equation}\label{eq1}
    \mc S_t = \frac{1}{N}\sum_{i,j=1}^N e^{- i t (E_i - E_j)}.
\end{equation}
This quantity can be easily generalized to account for finite temperature \cite{cotler_black_2017, cotler_chaos_2017, chenu_work_2019, martinez-azcona_analyticity_2022} or to include other filtering functions of the energies $g(E_i)$, e.g. with a Gaussian filter \cite{suntajs_quantum_2020}. 

In this article, we aim to decompose this quantity in terms of the spectral distance $k$, to understand the role played by different energy ranges---e.g. short-range neighbors with $k =1$ \textit{vs} long-range neighbors with $k \gg 1$---in building the universal shape of the SFF.  To do so, we introduce the contribution of the $k$th-energy level to the SFF, the $k$th neighbor SFFs, and study their properties. We then sum up each such contribution 
to obtain the complete SFF.  In this process, we learn about the underlying structure which can be unveiled via the decomposition of the SFF into $k$nSFF components.

\subsection{The ensemble averaged $k$th neighbor SFF}\label{sec:av_knSFF}

The SFF \eqref{eq1} can be rewritten as
\begin{equation}\label{SFF}
   \mathcal{S}_t  = \frac{N+2\sum_{i> j}\cos[t(E_i-E_j)]}{N^2} . 
\end{equation}
The numerator can be divided into terms according to spectral distances $s_i\ep k\equiv E_{i+k} - E_i$, which leads to the decomposition 
\begin{eqnarray} \label{SFF_decomposed}
    \mathcal{S}_t = \frac{1}{N}+\sum_{k=1}^{N-1} \mathcal{S}\ep k_t.
\end{eqnarray}
We thus consider the SFF as formed by summing the \textit{k-th neighbor spectral form factors} ($k$nSFF), that we introduce as 
\begin{eqnarray}
    \mathcal{S}\ep k_t \equiv \frac{2}{N^2} \sum_{i=1}^{N-k} \cos\left(t\,s_i^{(k)}\right) ~.
\end{eqnarray}

As briefly mentioned, the SFF is not self-averaging \cite{prange_spectral_1997}; so we wish to compute the ensemble average  
    $S_t  = 1/N+ \sum_{k=1}^{N-1} \langle \mathcal{S}\ep k_t \rangle$, 
which boils down to computing the ensemble average of the individual contributions from each spectral distance, denoted $\langle \mathcal{S}\ep k_t \rangle \equiv S_t\ep{k}$.  
We show below a series of approximations that allow us to derive an analytical expression for the $S\ep k_t$ which provides a good approximation.

We start from the definition 
\begin{align} \label{eq:knSFF1}
    &S_t\ep{k} = \frac{2}{N^2}\sum_{i=1}^{N-k} \int dE_1\dots dE_N \, \rho(\{E_i\})\, \cos[t\,s_i^{(k)}], 
\end{align}
where the energies $E_i$ are ordered. 
In the first step, we approximate all energy spacings $s\ep{k}_i=E_{i+k}-E_i$ as being independent of the absolute energy level\footnote{We expect such a behavior for an unfolded spectrum ($E_{i+k}-E_i$ does not depend on the local density of states $\rho(E_i)$) which shows no mixed behavior, i.e. the spectral statistics does not change within the energy window.} and set $i=1$ as the reference level, so  $s\ep k_i \rightarrow s\ep{k}_1$ and the sum over $i$ just becomes an $(N-k)$ factor. Then, we use the largest possible matrix that has a $k$-th level spacing and take $N\rightarrow(k+1)$. We recognize the distribution of the $k$-th level spacing, which can be approximated by the generalized Wigner surmise \eqref{Pk_dist}. With this, the $k$nSFF \eqref{eq:knSFF1} becomes 
\begin{align} 
    S_t\ep{k} &\approx \frac{2(N-k)}{N^2} \int ds\ep{k} \mathcal{P}\ep{k}(s\ep{k})\,\cos[t\,s\ep{k}],   
 \nonumber \\
        &\equiv  C_N^{(k)}\,f^{(k)}_t~, \label{Stk}
\end{align}
where, for later notational convenience, we have introduced the function and constant
\begin{subequations}
\begin{align}
    f^{(k)}_t &\equiv  \int ds \,\mathcal{P}\ep{k}(s)\,\cos(t\,s)\, , \label{ftk}\\
    C_N^{(k)} &\equiv \frac{2(N-k)}{N^2} \label{CNk}. 
\end{align}
\end{subequations}

Remarkably, the approach taken here can be generalized to quantities related to the SFF, such as autocorrelation functions. Indeed, the autocorrelation function at infinite temperature of a Hermitian operator $\ha O=\ha O^\dagger$ (i.e. any observable) can be decomposed in a similar manner using $f_t\ep{k}$ and changing only the coefficients $C_N\ep{k}$. To see this, consider the autocorrelation function 
\begin{eqnarray} \label{eq:correl_fct_def}
  \!\!  \mathcal{C}_t \equiv \frac{\mathrm{Tr}(\ha O \ha O(t))}{\mathrm{Tr}(\ha O^2)} =\frac{1}{\mathcal{N}^2} \!\!\sum_{i,j=1}^N |O_{ij}|^2 \cos\left[(E_{i}{-}E_j)t\right],
\end{eqnarray}
where $O_{ij}$ are the matrix elements of the operator $\ha O$ in the energy basis and $\mathcal{N}^2 = \sum_{i,j=1}^N |O_{ij}|^2$. Where $\ha O(t)$ denotes the time evolved operator $\ha O$ in the Heisenberg picture. 
 This expression can be decomposed according to level-spacing distances, namely 
\begin{equation}
\!\!    \mathcal{C}_t = \frac{1}{\mathcal{N}^2}\sum_{i=1}^N |O_{ii}|^2  {+} \frac{2}{\mathcal{N}^2}\!\! \sum_{k=1}^{N-1}\sum_{i=1}^{N-k} |O_{i,i+k}|^2 \cos(t s_i\ep{k}).
\end{equation}
Performing an ensemble average over the spectrum, which is taken as unfolded such that $E_{i+k}-E_i$ does not depend on $\rho(E_i)$, we can use the distribution $\mathcal{P}\ep{k}(s)$. The ensemble-averaged $\av{\mc C_t}$ follows as  
\begin{equation}
    \av{\mc C_t} = \frac{1}{\mathcal{N}^2}\sum_{i=1}^N |O_{ii}|^2 + \sum_{k=1}^{N-1} \av{\mc C\ep{k}_t} ,
\end{equation}
with the $k$th neighbor autocorrelation function 
\begin{eqnarray}
    \av{\mc C_t\ep{k}} \equiv O_N\ep{k} \, f_t\ep k.
\end{eqnarray}
Thus, knowledge of the function $f_t\ep{k}$, defined in \eqref{ftk} and whose analytical form is given below for the Gaussian and Poisson ensembles,  together with the coefficients $O_N\ep{k} \equiv \frac{2}{\mathcal{N}^2}\sum_{i=1}^{N-k} |O_{i,i+k}|^2$, provide the correlation function $\langle \mathrm{Tr}(\ha O \ha O(t)) / \mathrm{Tr}(\ha O^2) \rangle$.
%
The explicit relation between correlation functions and the $k$nSFF's is made explicit in App.~\ref{sec:protocol_knSFF}, together with a possible dissipative protocol to measure such correlation functions.

We now turn to derive analytical expressions for $f_t\ep{k}$, and thus $S_t\ep{k}$, for the RMT and Poisson ensembles.

\subsection{$k$nSFF in Gaussian random matrix ensembles} \label{sec:knSFF_Gaussian}
With the approximated analytical result for $\mathcal{P}\ep{k}(s)$, Eq.~(\ref{Pk_dist}), we can look for an expression of the $k$nSFF $S\ep{k}_t$ for the Gaussian random matrix ensembles. Starting from (\ref{Stk}), we have 
\begin{eqnarray}
   \!\!\!  f_t^{(k)} \equiv \! \int_0^\infty \!\!\! d s \, \mathcal{P}\ep{k}(s)\cos(t s) = \sum_{n=0}^\infty \frac{({-}1)^n}{(2n)!}t^{2n}  \langle s^{2n} \rangle
\end{eqnarray}
with
\begin{align}
    \langle s^{2n} \rangle &= \int_0^\infty d s \, s^{2n} \,\mathcal{P}\ep{k}(s) = \left(\frac{\alpha+1}{2}\right)_n (A_\alpha)^{-n}
\end{align}
where we have used the Pochhammer symbol $(a)_n = a(a+1)\dots(a+n-1)=\Gamma(a+n)/\Gamma(a)$. 
The function becomes
\begin{eqnarray}
    f_t^{(k)} &=&  \sum_{n=0}^\infty \frac{\left(\frac{\alpha+1}{2}\right)_n}{(1/2)_n}\, \frac{\left(-t^2/4 A_\alpha  \right)^n}{n!} ~ ,
\end{eqnarray}
where the sum can be expressed in terms of a hypergeometric function, namely 
\begin{eqnarray} 
    f_t^{(k)} &=& e^{- \frac{t^2}{4 A_\alpha}}\,{}_1F_1 \left(-\frac{\alpha}{2}; \frac{1}{2}; \frac{t^2}{4 A_\alpha}\right).
\end{eqnarray}
Since the coefficient $1/(2\sqrt{A_\alpha})$ appears in the $t$-dependent exponent, it is homogeneous to a frequency $\omega_k$. We set 
\begin{equation}\label{eq:frequency}
\frac{1}{2 A_\alpha} \equiv \frac{\omega^2_k}{\alpha}; 
\end{equation}
we will see that this quantity sets the width of the Gaussian envelope. 
The hypergeometric function  can itself be expressed in terms of a Laguerre function, so we get
\begin{eqnarray} \label{knSFF_Laguerre_final}
    f_t^{(k)} 
    &=& \sqrt{\frac{\pi \alpha}{2}}\frac{k}{\omega_k} e^{- \frac{\omega_k^2 t^2}{2 \alpha}} L_{\alpha/2}^{-1/2}\left( \frac{\omega_k^2t^2}{2 \alpha} \right).~ \label{ft_Lag}
\end{eqnarray}
The Laguerre \textit{function}, $L_{\mu}^{a}$ with $m-1< \mu < m$ for $m\in \mathbb{N}$ and $a >-1$, is defined by the infinite sum  \cite{mirevski_fractional_2010} 
\begin{eqnarray} \label{Laguerre_nonint_def}
    L_\mu^a(z) &=& \sum_{k=0}^\infty \binom{\mu+a}{\mu-k} \frac{(-z)^k}{k!} ~.
\end{eqnarray}
When $\alpha$ is even, $\alpha=2m$ for $m\in \mathbb{N}$, the Laguerre function in (\ref{knSFF_Laguerre_final}) becomes a Laguerre \textit{polynomial} of degree $m$:
\begin{eqnarray} \label{Laguerre_def}
    L_m^a(z) &=& \sum_{k=0}^m \binom{m+a}{m-k} \frac{(-z)^k}{k!}~.
\end{eqnarray}
Since the degree of the Laguerre polynomial (or function, for non integer $\alpha/2$) grows quadratically with $k$, we can use the  approximation for Laguerre polynomial of high degree \footnote{See Digital Library of Mathematical Functions \url{https://dlmf.nist.gov/18.15\#iv}}:
\begin{eqnarray}
    L_n^a(x) &=& \frac{n^{a/2-1/4}}{\sqrt{\pi} x^{a/2+1/4}} e^{x/2}  \Big[\cos(\theta_{a,n}(x)  )\left(  1+ \mc{O}\left(\frac{1}{n}\right)\right) \nonumber\\
    &&+ \sin(\theta_{a,n}(x) )\left(  \frac{b_a(x)}{\sqrt{n}}+ \mc{O}\left(\frac{1}{n}\right)\right)\Big] \label{Ln_approx}
\end{eqnarray}
where $ \theta_{a,n}(x) = 2\sqrt{n x} -a\pi/2-\pi/4$ and $b_a(x) = \big(4x^2-12a^2-24ax-24x+3\big)/(48 \sqrt{x})$. 
Note that  $b_{-1/2}(x)=\frac{\sqrt{x}}{12}(x-3)$.

\begin{figure*}
    \centering
    \includegraphics[width = \textwidth]{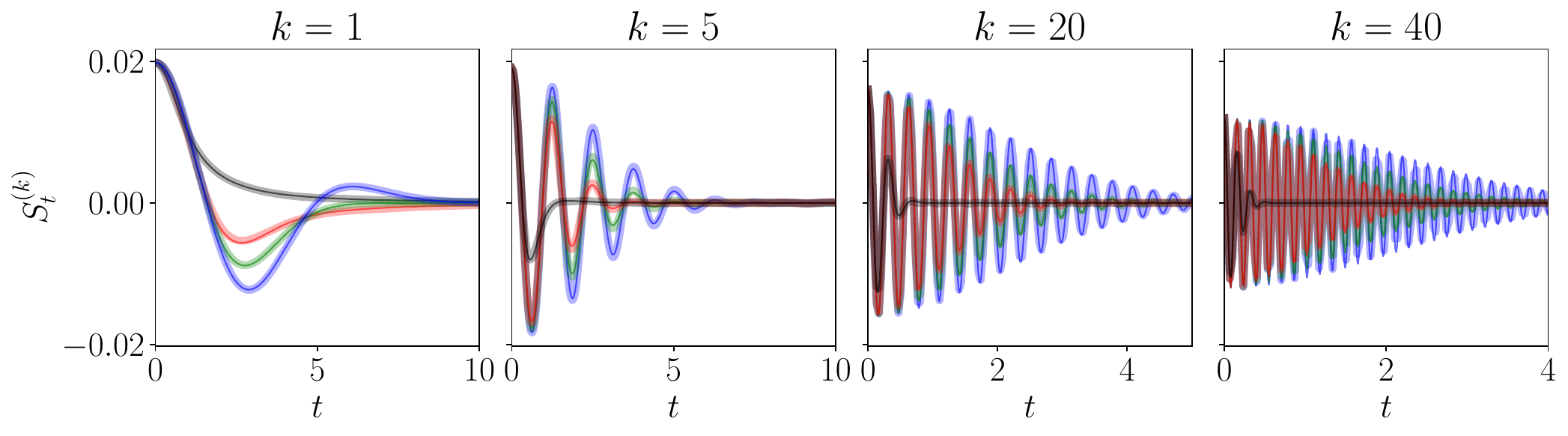}
    \caption{\textbf{Time evolution of the $k$nSFF} for Poisson (black), GOE (red), GUE (green) and GSE (blue) for different spectral neighbors, $k = 1, 5, 20, 40$ in systems of dimension $N=100$. The plots for $k=5, \; 20, \; 40$ show analytical results (thin lines) from \eqref{SFFk_Pois} and \eqref{SFFk_Approx}, while for $k=1$ we show the exact expression \eqref{knSFF_Laguerre_final}, and numerical results for random matrices averaged over $N_\mr{av}=1000$ realizations (thick transparent lines). While we do not expect the approximation \eqref{SFFk_Approx} to be good for small $k$, it works already quite well for GSE and $k=1$ and less so for GUE and GOE, in that order. Note the different scales in the time axis, chosen to better represent the increasing number of oscillations with the spectral neighbor $k$, see Eq.~\eqref{eq:nboscillations}.}
    \label{fig:SFFk}
\end{figure*}

In our case, we have a Laguerre polynomial of degree $n=\alpha/2$ (not necessarily integer), $a=-1/2$ and $x=\frac{\omega_k^2 t^2}{2 \alpha}$, so the above approximation reads
\begin{eqnarray}
 \!\!   L_{\frac{\alpha}{2}}^{-\frac{1}{2}}(x) \approx \sqrt{\frac{2}{\pi \alpha}}e^{\frac{x}{2}}\Big[ \cos( \omega_k t)
    +\frac{b_{-\frac{1}{2}}(x)}{\sqrt{\alpha/2}}\sin( \omega_k t ) \Big]. \label{Lag_approx}
\end{eqnarray} 
As can be seen in (\ref{Ln_approx}), the terms in the square brackets of (\ref{Lag_approx}) are an approximation up to (not including) $\mc O(1/\alpha)$.  Thus, when using this result into (\ref{ft_Lag}), we must ensure that the combined coefficient of the square brackets is expanded to the same order in $\alpha$, 
\begin{eqnarray}
    \frac{1}{\sqrt{\alpha/2}}\frac{\Gamma(\alpha/2+1)}{\Gamma(\alpha/2+1/2)} = 1 +\mc{O}(1/\alpha) ~.
\end{eqnarray}
Eventually, we find that $f_t\ep{k}$ can be approximated by 
\begin{eqnarray}\label{approxftk}
    f_t\ep{k} &=&  \Big(1+\mc{O}(1/\alpha)\Big) e^{-\frac{\omega_k^2 t^2}{4 \alpha}} \\
    &\times&\Big[ \cos (\omega_k t)+\sqrt{\frac{2}{\alpha}}{b_{-\frac{1}{2}}\left(\frac{\omega_k^2 t^2}{2 \alpha}\right)}\sin(\omega_k t)+\mc{O}(1/\alpha)\Big].\nonumber
\end{eqnarray}
Note that we are not expanding terms involving time in $\alpha$. 
We note that the frequency 
is well approximated at large $k$ by a linear function in $k$:
\begin{eqnarray}\label{app_w_k}
    \omega_k 
    \approx k -\frac{1}{2\beta k}+ \mc{O}(1/k^2)~.
\end{eqnarray}

Let us make several remarks about expression (\ref{approxftk}):
\begin{itemize}
    \item The initial value is always equal to unity, $f_{t=0}\ep{k}=1$ for all $k$;
    \item For $t\to \infty$, the overall exponential factor, $e^{-\frac{\omega_k^2 t^2}{4 \alpha}}$, makes $f_t\ep{k} \to 0$;
    \item $f_t\ep{k}$ is expressed as a sum of a cosine and a sine with the same frequency $\omega_k \to k$ at large $k$;
    \item Apart from the overall exponential factor, the coefficient of $\cos(\omega_k t)$ is $1$ while the coefficient of $\sin(\omega_k t)$ is time-dependent and is equal to $\sqrt{\frac{2}{\alpha}}\,b_{-1/2}\left(\frac{\omega_k^2 t^2}{2 \alpha}\right)=\frac{1}{12 \alpha} \omega_k t \left(\frac{\omega_k^2 t^2}{2 \alpha}-3\right)$. It is of $\mc O(1/k)$ and thus of less consequence for large $k$. Note that, at the same time, it is more significant at large $t$. This is compatible with the fact that small $k$ terms (corresponding to low frequencies) are more significant at long time scales.
    \item We can compute the number of oscillations in one standard deviation of the envelope $\sqrt{2 \alpha} /\omega_k$, by  comparing it with the period of the oscillations $T_k = 2 \pi/\omega_k$. We thus find
    \begin{equation}\label{eq:nboscillations}
        \frac{\sqrt{2 \alpha}}{2 \pi } \xrightarrow[k\to\infty]{} \frac{\sqrt{\beta}}{2 \pi} k.
    \end{equation}
    So the number of oscillations of the $k$nSFF in the envelope is proportional to $\sqrt{\alpha}$,  and scales linearly with the neighbor degree $k$ for large $k$, as illustrated in Fig. \ref{fig:SFFk}. The figure also illustrates that the largest number of oscillations that happen before the signal flattens because of the exponential decays is for the GSE, which has the largest $\beta$;
    \item Were the $k$nLS distribution a perfect Gaussian centered at $\av{s\ep k} =k$, the $k$nSFF would only involve the Gaussian envelope and the cosine term $f_t\ep k = e^{-k^2 t^2/(4 \alpha)}\cos (k t)$ with frequency $\omega_k = k$, since the Fourier transform of a Gaussian is another Gaussian. The non-zero mean is accounted for by including $e^{i \av{s\ep k} t}$, whose real part is $\cos(kt)$. Thus the sine term in the $k$nSFF comes from the non-Gaussianity of the $k$nLS distribution. 
\end{itemize}

With the approximation (\ref{approxftk}), the final expression for $S_t\ep{k}$ becomes 
\begin{eqnarray}\label{SFFk_Approx}
    S_t\ep{k} &\approx& \frac{2(N-k)}{N^2} e^{-\frac{\omega_k^2 t^2}{4 \alpha}} \Big[ \cos (\omega_k t)\\
    &&+\frac{1}{12 \alpha} \omega_k t\Big(\frac{\omega_k^2 t^2}{2 \alpha}-3\Big)\sin(\omega_k t)\Big]. \nonumber
\end{eqnarray}
 Figure \ref{fig:SFFk} shows that our analytical approximation above reproduces well the numerical data for all three Gaussian ensembles.
 

\subsection{$k$nSFF for Poisson ensemble} \label{sec:knSFF_Pois}
The averaged $k$th neighbor SFF can be computed for matrices with uncorrelated eigenenergies, in the same manner as for the Gaussian ensembles case, but now using the corresponding probability distribution, Eq. \eqref{Pk_Pois}. We thus find
 \begin{equation}
 \label{SFFk_Pois}
     S_t\ep{k}=\frac{2(N-k)}{N^2}\frac{\cos(k \arctan t)}{(1+t^2)^{k/2}} ~.
 \end{equation}

This expression shows a Lorentzian envelope multiplied by a cosine function; note that the cosine function gives oscillations, but they are not periodic due to the $\arctan(t)$ in the argument. Fig.~\ref{fig:SFFk} (black curves) shows that this expression captures well the numerical results, and that the envelope for Poisson has a much smaller width, suppresing the oscillations much faster than for the RMT ensembles.


\section{The properties of the $k$-th neighbor Spectral Form Factors}\label{sec:properties_knSFF}

Before turning to the complete SFF, we study some more properties of the $k$th neighbor SFF, in particular, the time it reaches its minimum, $t_m$ and the corresponding depth as a function of $k$. We also look at the scaling of the neighbor $k^*$ for which the $k$nSFF is the deepest, as a function of the system's dimension.

\subsection{Minimal value of $k$nSFF and associated time}
We now study the time at which each $S_t\ep{k}$ reaches its minimal value for both RMT and Poisson ensembles.
\subsubsection{Random Matrices}
For Random matrices, the `minimum time' $t_m(k)$ can be computed from the exact expression using $f_t\ep{k}$ given by \eqref{knSFF_Laguerre_final}, or from the approximate one, Eq. \eqref{SFFk_Approx}. However it is challenging to analytically determine the minimum of those functions. One possibility to overcome this is take \eqref{SFFk_Approx} for large enough $k$ (and not too large $t$). Then, the main contribution comes from the cosine. We know that its minimum happens when its argument is equal to $\pi$, therefore 
\begin{equation}\label{approx_td}
    t_m(k) \approx \frac{\pi}{\omega_k} \stackrel{\eqref{app_w_k}}{\approx} \frac{\pi}{k}. 
\end{equation}
Note that this result implies that the minimum time does not depend on the ensemble $\beta$, but only on the neighbor degree $k$. Fig.~\ref{fig:comparison_num_rmt}(a) confirms this observation as the behavior predicted by \eqref{approx_td} very well captures results for the three RMT ensembles, shown in the dots.
From this estimate, we can also find the $k$nSFF depth as a function of $k$. Inserting \eqref{approx_td} into \eqref{SFFk_Approx}, we obtain
\begin{eqnarray}
\label{eq:minStd_RMT} 
    S_{t_m}\ep{k} 
    &\sim& -\frac{2(N{-}k)}{N^2} e^{-\frac{\pi^2}{2 k(\beta k + \beta + 2)}}.
\end{eqnarray}
Note that this expression has two contributions: the prefactor comes from the number of $k$nLS, while the exponential depends on the Dyson parameter $\beta$. This expression is plotted in Fig.~\ref{fig:St_dip}(left), which shows that at small $k$, the minimum values obtained by different $k$nSFF's depend on the ensemble, with deeper minima for larger $\beta$. In turn, around $k \approx 10$ the three ensembles converge to a similar curve. Furthermore, the analytical results (dashed lines) are in very good agreement with the numerical RMT results (solid lines).

\subsubsection{Poisson Ensemble}
\begin{figure}
    \centering
    \includegraphics[width = \linewidth]{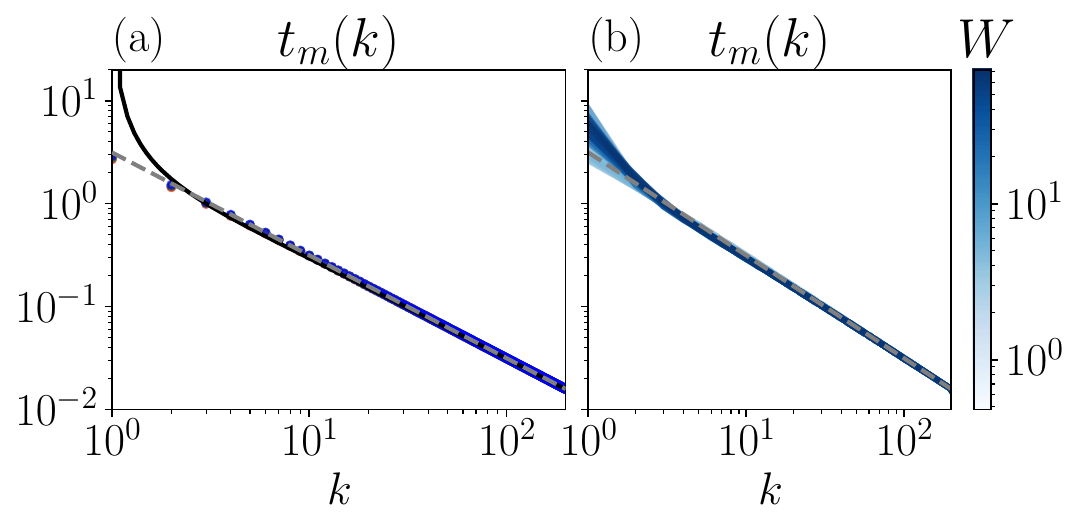}
    \caption{\textbf{Minimum time} as a function of the neighbor degree for (a) ideal ensembles: Poisson (black), GOE (red), GUE (green) and GSE (blue) computed numerically from the unfolded spectrum along with the approximate expression \eqref{approx_td} (dashed grey). (b) Results for the XXZ spin chain for different values of the disorder (blue colorscale) along with the analytical result $t_m(k) = \pi/k$ (dashed gray).
    }
    \label{fig:comparison_num_rmt}
\end{figure}

In the Poisson ensemble, the minimum time is the first minima of the function \eqref{SFFk_Pois}. It can be computed exactly as  
\begin{eqnarray}\label{tmin_k_Pois}
    t_m(k) = \tan\left(\frac{\pi }{1+k}\right)~.
\end{eqnarray}
Note that it \pablo{diverges at $k=1$ because the first $k$nSFF for the Poissonian ensemble shows no dip and} asymptotically goes as $\pi/k$, similarly to the Gaussian ensembles. Figure \ref{fig:comparison_num_rmt} (a) shows that the minimum time for Poisson \eqref{tmin_k_Pois} fastly converges to the RMT approximation \eqref{approx_td}. This suggests that the period of oscillations of the $k$nSFF cannot be used to differentiate between RMT and Poisson for $k\geq 2$. In other words, this minimum time does not capture (except for $k=1$) on the presence or absence of chaos in the system. 
Furthermore, the minimum time agrees with the numerical simulation from the different Gaussian ensembles and for the XXZ model (which is introduced in Sec. \ref{sec:XXZ}), provided that the spectrum is unfolded.

The minimum value of $\av{S_t\ep{k}}$  is 
\begin{eqnarray}\label{eq:Std_Poi}
    S_{t_m}\ep{k} = \frac{2(N-k)}{N^2} \cos^k\left(\frac{\pi}{1+k}\right)\cos\left(\frac{k\pi}{k+1}\right). 
\end{eqnarray}
Figure~\ref{fig:St_dip}(left) shows that this analytical expression (dashed black) agrees with numerical realizations for the Poissonian ensemble. At $k=1$, the $k$nSFF in Poisson shows a minima at zero. Perhaps counterintuitively, the $k$nSFF's for Poisson can have a quite deep minima for large $k$, with its minimum being deeper than the one shown for $k=1$ for GOE and GUE. Interestingly, at very large spectral distance, over $k \sim 100$, the minima of the SFF in Poisson converges to the minima in the RMT ensembles. This is the regime in which the prefactor $-2(N-k)/N^2$ dominates in both expressions \eqref{eq:Std_Poi} and \eqref{eq:minStd_RMT}. This occurs because both of the remaining functions in the expressions converge to unity at long times. This convergence happens with terms of $\mc O(k^{-2})$ for RMT and of $\mc O(k^{-1})$ for Poisson, which explains the faster convergence of random matrices. In this regime, the minimum of the $k$nSFF's is made shallower due to the small number of neighbors at long range.

\subsection{The deepest $k$-nSFF}
We now look at the spectral distance $k^*$ associated to the $k$nSFF with the deepest minimum, and see if it can capture some aspects of energy correlations. \pablo{This value represents the spectral distance $k^*$ such that the $k$nSFFs with shorter range ($k<k^*$) get deeper, while those with longer ranger ($k>k^*$) get shallower}. 
It can be computed from the minima of Eq.~\eqref{eq:minStd_RMT} for the Gaussian random matrix ensembles and \eqref{eq:Std_Poi} for the Poissonian ensemble.

\subsubsection{Random Matrices}
For Gaussian random matrices, it is challenging to extract the minimum from the exact expression due to the interplay between the oscillations and the envelope. Taking the large $k$ approximation, we find the deepest $k^*$ to be given approximately by the solutions of
\begin{equation}\label{k_st_rmt}
    k^{*2}(2 + \beta + \beta k^*) = N \pi^2,
\end{equation}
whose exact solution can be found exactly using \textit{Mathematica} but which is too cumbersome to bring any insight. However, a large $N$ expansion  yields
\begin{equation}\label{eq:kstar}
 k^* \sim \mc C_\frac{1}{3} N^\frac{1}{3} + \mc C_0 + \mc C_{-\frac{1}{3}} N^{-\frac{1}{3}} + \mc O(N^{-\frac{2}{3}})
 \end{equation} 
 for the Gaussian ensembles, where the coefficients are given by $\mc C_\frac{1}{3}=\left(\frac{\pi^2}{\beta}\right)^\frac{1}{3}$, $\mc C_0 = - \frac{\beta + 2}{3 \beta} $, $\mc C_{-\frac{1}{3}}= \frac{(2 + \beta)^2 - 3 \beta \pi^2}{9 \beta^\frac{5}{3}\pi^\frac{2}{3}}$. Thus, the spectral range with the deepest $k$nSFF scales with the cube root of the dimension of the system. Fig.~\ref{fig:St_dip}(right) shows a good agreement between this expression for different values of $\beta$ and numerical random matrices for the three Gaussian ensembles, specially at large $k$. It also shows that the deepest $k$nSFF is reached first in the GSE, then in the GUE and lastly in the GOE.

 \subsubsection{Poisson Ensemble}
\begin{figure}
    \centering
    \includegraphics[width=\linewidth]{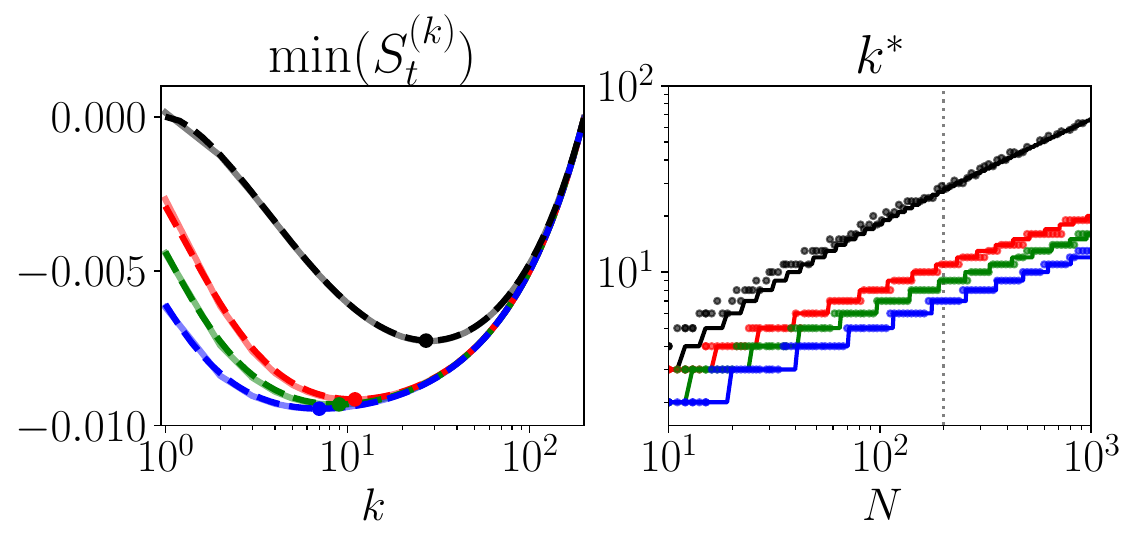}
    \caption{ (left) \textbf{Minimum value of $k$nSFF $S_t\ep{k}$ as a function of the spectral distance $k$}. Approximate analytical results (dashed lines) for the RMT ensembles \eqref{eq:minStd_RMT} and Poisson distribution \eqref{eq:Std_Poi}, and numerical results (solid lines). Colors are as in Fig.~\ref{fig:SFFk}, i.e. Poisson (black), GOE (red), GUE (green) and GSE (blue). The  numerical results are obtained from matrices of dimension $N=200$ and averaged over $N_\mr{av}=2000$ elements of the ensemble. Note that the function at large $k$ is linear, even if the choice of logarithmic scale in the $k$ axis does not allow a simple visualization. (right) \textbf{Scaling of the deepest neighbor $k^*$} as a function of the system size $N$ computed numerically for RMT and Poisson ensembles (circles) and the analytical approximations \eqref{eq:kstar} and \eqref{eq:kstar_poi} rounded to the nearest integer (lines). Numerical results are averaged over $N_\mr{av}=200$ matrices. We show a guide for the eye at $N=200$ (gray dotted line) which agree with the values of $k^*$ used in Figure~\ref{fig:XXZ_Stdip} for the Poissonian and GOE endpoints of $W$. }
    \label{fig:St_dip}
\end{figure}

 The expression for the minimum of the $k$nSFF for the Poisson ensemble \eqref{eq:Std_Poi} admits the asymptotic expansion
\begin{equation}
    S_{t_m}\ep k = \frac{2 (N-k)}{N^2}\left(- 1 + \frac{\pi^2}{2 k}- \frac{4 \pi^2 + \pi^4}{8 k^2} + \mc O(k^{-3})\right),
\end{equation}
which leads to a third-order polynomial equation whose solution can be expanded for large $N$. We thus find the deepest $k$nSFF to scale as
\begin{equation}\label{eq:kstar_poi}
    k^*_0 \sim \frac{\pi}{\sqrt{2}}N^\frac{1}{2} - \left( 1 + \frac{\pi^2}{4}\right)+\mc O(N^{-\frac{1}{2}}), 
\end{equation}
which scales faster than for the Gaussian ensembles \eqref{eq:kstar}, i.e. with the square root of the dimension of the system. Thus, the deepest $k$nSFF in the Poissonian ensemble happens for  $k^*$ larger than in the chaotic case, as seen in Fig.~\ref{fig:St_dip}(right). The Figure also shows good agreement of the above analytical approximation with the numerics, specially at large $k$. 

To end this section, we recall that the total SFF is not self-averaging. Then, it will not come as a surprise that the $k$nSFF is also not a self-averaging quantity. For more details and results in the GUE, see App. \ref{app:self-av}.


\section{Building the SFF from the $k$-nSFF's}
\label{sec:building_SFF_fromknSFF}

In this section, we study how the decomposition of the Spectral Form Factor in terms of $k$nSFF's can help to understand the build-up of the correlation hole. We begin by studying the partial SFF with neighbors only in the range $k \in [1, \dots, K]$ and studying the characteristic time-scales of that quantity, namely the \textit{dip} and \textit{Thouless} time. We then report an intriguing result showing the contributions from even and odd neighbors, and finally discuss the full SFF as obtained  from the $k$nSFF's.

\subsection{Building the correlation hole: the partial SFF}
We now have the tools to ask how the correlation hole and ramp build up (or not) in different systems. This can be answered by introducing the  \textit{partial} SFF, which we define as 
\begin{align} \label{SFF_Kmax}
    S_{t,K} \equiv \frac{1}{N}+\sum_{k=1}^{K}S_t\ep{k}~,
\end{align}
and represents the SFF with a cut-off $K$ on the neighbor range contributing to it. Note that $S_{t,N-1}=S_t$.
When it exists, we observe that the ramp progressively builds up by adding up contributions $S_t\ep{k}$ with longer range $k$, as detailed below.

\subsubsection{Time-scales of the ramp: Dip and Thouless times}

To characterize how the ramp is built, we introduce two time-scales. First, we define the \textit{partial} dip time as the time after which the initial oscillations of the SFF stop being important and the SFF starts growing on average. This increase is not monotonous, due to SFF fluctuations originating from quantum noise. So we propose to look at the absolute minima of the relative maxima of the partial SFF, specifically:
\begin{itemize}
    \item Compute the set of relative maxima of the partial SFF $t_\text{rel-max}$, i.e. $t_\text{rel-max}:=\{t \in \mathbb R, \; \mr{s.t.}\; \partial_t S_{t, K}|_t=0, $$\; \partial_t^2 S_{t, K}|_t<0$\};
    \item Given this set of times, compute the absolute minima of the relative maxima, i.e. $t_\mr{dip}$ such that $S_{t_\mr{dip}, K}= \min S_{t_\text{rel-max}, K}$. 
\end{itemize}
This definition ensures that after the dip time, the relative maxima (including the ones coming from quantum noise) grow, such that this time can be interpreted as the onset of the SFF growth,  i.e. the beginning of the ramp. 

\begin{figure}
    \centering
    \includegraphics[width=\linewidth]{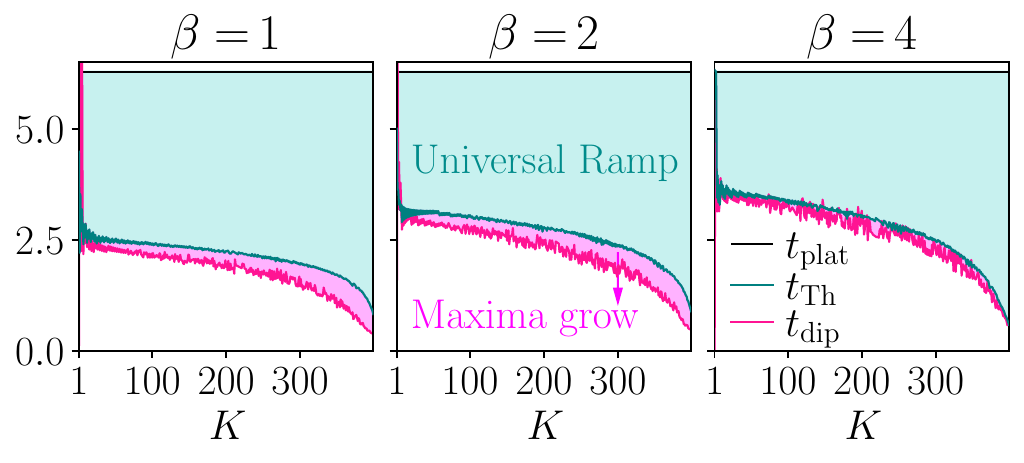}
    \caption{Time scales for the partial $K$-neighbors SFF. The plots show the dip (pink), Thouless (turquoise), and plateau $t_p=2 \pi$ (black) times as a function of the maximum number of neighbors $K$ considered, for the three Gaussian ensembles. The shaded regions represent the part where the SFF grows in a non-universal way (pink) and where it grows with the universal ramp of the connected SFF (light blue).  The results are computed with $N_\mr{en}=400$ from the analytical expressions of the SFF for Random Matrices \eqref{SFFk_Approx}. For the Thouless time, we used the partial SFF without smoothing, taking $\epsilon = 0.1$ for GOE and GUE and $\epsilon = 0. 25$ for GSE. This choice is due to the challenge in building the full spike of the GSE from summing $k$nSFF's (cf. Apps. \ref{app:dip_time_SFF}, \ref{App:SFF_test}).}
    \label{fig:tdip}
\end{figure}

The second time scale that is natural to define is the \textit{Thouless time}, originally introduced in a single-particle context \cite{Edwards_1972}. In the particular case of the SFF, we follow Suntajs \textit{et al.} \cite{suntajs_quantum_2020} and define it through the logarithmic ratio between the partial and the connected SFF's
\begin{equation}\label{eq:def_DeltaStk}
    \Delta S_{t, K} = \left|\log_{10}\left(\frac{S_{t, K}}{b_t}\right)\right|,
\end{equation}
where $b_t$ is the universal connected part of the SFF. The choice between GOE, GUE or GSE is done according to the symmetries of the system. The Thouless time is then defined as the time such that $\Delta S_{t_\mr{Th}, K} = \epsilon$ and $\Delta S_{t, K}< \epsilon$ $\forall t > t_\mr{Th}$. This time thus captures the time after which, considering only up to $K$ neighbor correlations in the SFF, we follow the universal ramp of the connected SFF. 
Further explanation of the meaning of both times are given in App. \ref{app:dip_time_SFF}. 

The partial SFF is not necessarily positive if $K< N-1$, which means that some of the oscillations become negative. This poses a problem for the smoothing of the SFF, since the quantity has very pronounced oscillations. For this reason, contrary to the approach followed in \cite{suntajs_quantum_2020}, we decided not to smoothen the partial SFF for the computation of these time scales. Note that taking an absolute value of the partial SFF would bring extra maxima to the quantity, and thus does not solve this issue.

\begin{figure*}
    \centering
    \includegraphics[width = \linewidth]{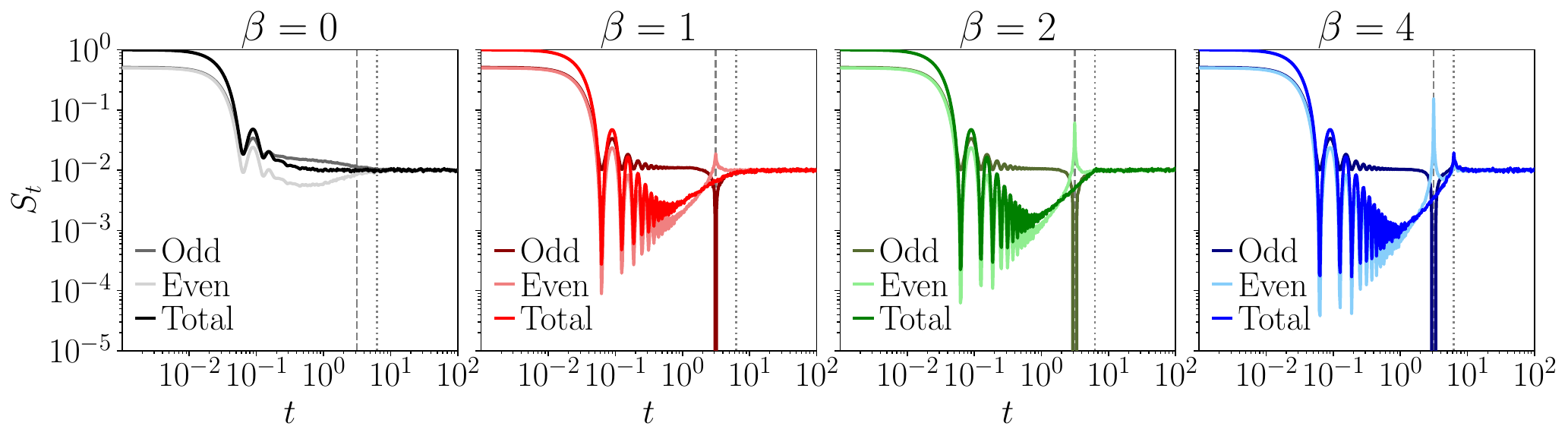}
    \caption{\textbf{Odd vs even neighbor contributions to the SFF and their sum for Poisson, GOE, GUE and GSE}; computed numerically from $N_\mathrm{av}=1000$ matrices of dimension $N=100$. For visualization of the data in log-log scale, an extra factor of $1/2N$ was added to the even and odd contributions. The even contributions construct a `resonance' while the odd ones construct an `anti-resonance'.
    The  vertical lines highlight the time at which the resonance and  anti-resonance happen, $t^* = \pi$ (dashed gray), and at which the plateau starts for the Gaussian ensembles, $t_p = 2 \pi$ (dotted gray).}
    \label{fig:even_vs_odd}
\end{figure*}

The results obtained for these time-scales are shown in Fig. \ref{fig:tdip}. The partial SFF has been computed from the analytical approximations for the different RMT ensembles. We see that both the dip and Thouless times decrease as we increase the number of neighbors, i.e. the ramp gets longer. Interestingly, we also observe here that the dip time precedes the Thouless time, this means that the relative maxima of the partial SFF start to grow before the partial SFF gets close to the connected SFF. Both of these time-scales also show that adding the intermediate neighbors does not decrease too much the dip and Thouless times, suggesting that the neighbors that contribute the most to the ``duration'' of the ramp are the short-range neighbors $k<5$ and the very long-range neighbors $k>150$ for $N=200$. From these observations, we conclude that all neighbors are needed to explain the full duration of the ramp, although some contribute more than others.

A similar argument can be made from the relationship between the two-level correlation function $R_2(s)=\sum_{k=1}^\infty \mathcal{P}\ep{k}(s)$ and the probability densities $\mathcal{P}\ep{k}(s)$ given in \cite{pandey2019quantum}, as discussed in App.~\ref{App:SFF_connected}. From here we find that small, medium, and large $k$ neighbor correlations contribute to the structure of the connected SFF.

\subsection{The contribution from even and odd neighbors}\label{sec:even_odd}

\begin{figure*}
    \centering
    \includegraphics[width = \linewidth]{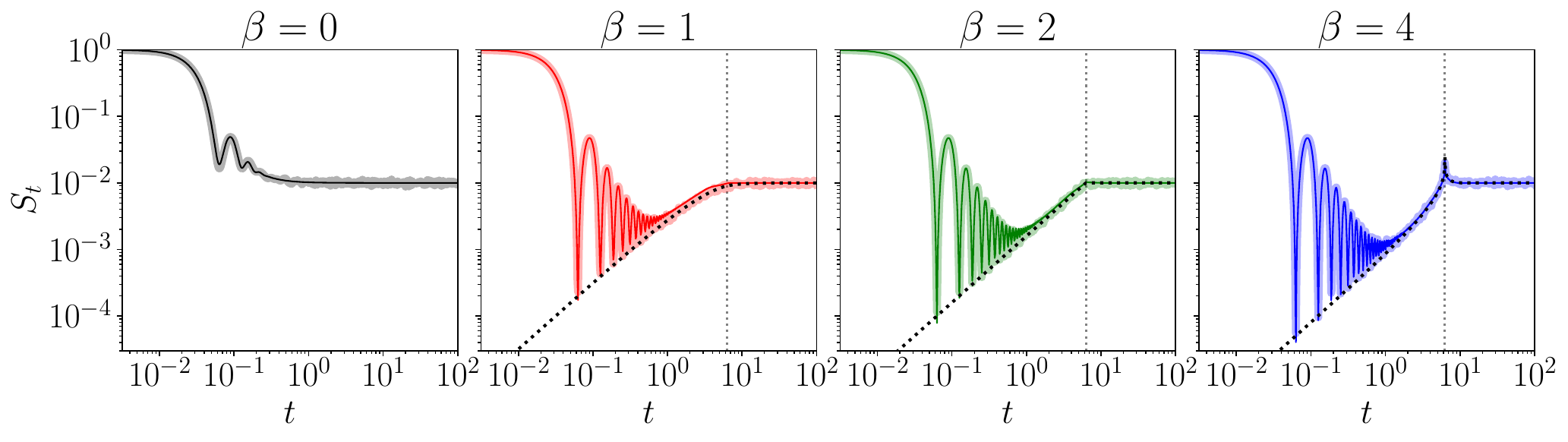}
    \caption{\textbf{Spectral Form Factor} for: Poisson (black), GOE (red), GUE (green) and GSE (blue) computed numerically (thick transparent line) and using the analytical results (thin solid line) given by \eqref{SFF_Pois} (black) and \eqref{SFF_approx} (red, green, blue). The connected SFF for each of the ensembles, see App. \ref{App:SFF_connected}, is also shown (black dotted line). The plots show results for random matrices with dimension $N = 100$ and the numerics have been averaged over $N_\mathrm{av}=1000$ matrices. The dotted gray line marks the start of the plateau at $t_p = 2 \pi$. }
    \label{fig:SFF_approx}
\end{figure*}

The approximation (\ref{SFFk_Approx}) for the ensemble average of the $k$nSFF's for RMT is expressed as a combination of cosines and sines with frequency $\omega_k$ given by (\ref{app_w_k}). 
We wondered about the even and odd range contributions, defined as
\begin{subequations}
\begin{align}
    \langle \mathcal{S}_t^{(\text{even})}\rangle &\equiv \frac{1}{2N}+ \sum_{k \; \mr{even}} \av{\mathcal{S}_t\ep{k}} \\
    \langle \mathcal{S}_t^{(\text{odd})}\rangle &\equiv \frac{1}{2N}+\sum_{k \; \mr{odd}} \av{\mathcal{S}_t\ep{k}}.
\end{align}  
\end{subequations}
These are presented in Fig.~\ref{fig:even_vs_odd} for random matrices taken from GOE, GUE, and GSE ensembles of dimension $N=100$.
Inspecting the plots, we find a constructive interference for $\langle \mathcal{S}_t^{(\text{even,odd})}\rangle$ at $t\approx\pi$. More specifically, there is a ``resonance" (positive  peak) for $\langle \mathcal{S}_t^{(\text{even})}\rangle$ and an ``anti-resonance" (negative peak) for $\langle \mathcal{S}_t^{(\text{odd})}\rangle$ at $t\approx \pi$. This observation can be explained by our analytical results,
\begin{align}
     S_t^{(\text{even/odd})} &\approx \frac{1}{2N}+ \sum_{k \text{ even/odd}} C_N\ep{k} f_t\ep{k} \label{Steven}
     \end{align}
where $C_N\ep{k}$ is given by (\ref{CNk}) and $f_t\ep{k}$ is given by the approximation (\ref{approxftk}). Taking $\omega_k\approx k$, the sum of even neighbor ranges involves a sum of $\cos(k\,t)$ with $k=2,4,6,\dots$ which interfere constructively at $t=\pi$ to create a ``resonance"; similarly, the sum of odd neighbor ranges involves a sum of $\cos(k\,t)$ with $k=1,3,5,\dots$ which interfere constructively at $t=\pi$ to create an ``anti-resonance".  \pablo{Interestingly, Fig. \ref{fig:even_vs_odd} also shows that most of the ramp is constructed from the even neighbors contribution, while the odd terms mostly contribute to cancel the even resonance and render the universal ramp.}  
We also note that the `spike' seen in the complete SFF for GSE (as can be seen in Fig. \ref{fig:SFF_approx}) is nothing but the next constructive interference from both the even and odd contributions, and happens (for the unfolded spectrum) at $ t=2\pi$. Note that the transition from the ramp to the plateau also happens at $t_p=2 \pi$ for GUE and GOE.


\subsection{The full SFF}

We now add up the contributions from all spectral distances to write the total SFF and compare our approximate analytical results to numerical simulations. Using the approximation \eqref{SFFk_Approx} for the $k$th neighbor SFF, the total averaged SFF for the Gaussian ensembles follow as
\begin{align} \label{SFF_approx}
    S_t \approx \frac{1}{N}+\sum_{k=1}^{N-1}&\frac{2(N-k)}{N^2} e^{-\frac{\omega_k^2 t^2}{4 \alpha}} \Big[ \cos (\omega_k t)\\
    &+\frac{1}{12 \alpha} \omega_k t\Big(\frac{\omega_k^2 t^2}{2 \alpha}-3\Big)\sin(\omega_k t)\Big]. \nonumber
\end{align}
Figure \ref{fig:SFF_approx} present a comparison of our analytical results with numerical results for the random matrix ensembles. It shows that for the Gaussian ensembles, the above approximate expression gives good results, even without using any exact results for $S_t\ep{k}$. In particular, the transition between the ramp and plateau is well captured for the three ensembles: smooth for GOE, `kink' for GUE and `spike' for GSE. The time at which this transition happens was first discussed in \cite{alhassid_spectral_1992} and we provide an alternative rationale for it by decomposing the SFF into the contributions from odd and even spectral distances, as discussed in the previous section.  Importantly, all the ensembles, when unfolded, show the plateau time at $t_p = 2 \pi$.  This is consistent with the results in \cite{alhassid_spectral_1992} which show the plateau for the Gaussian ensembles at $t_p/(2 \pi \bar \rho)=1$. In our case unfolding the spectra sets $\bar \rho = 1$ and thus $t_p = 2 \pi$. 
A careful observation of Fig. \ref{fig:SFF_approx} shows that the SFF computed from our analytical expressions slightly overestimates the actual value of the ramp, specially for GOE. In Appendix \ref{App:SFF_test}, we test the accuracy of the total result given in Eq. \eqref{SFF_approx} with respect to the SFF for RMT computed numerically and with respect to the connected SFF.

One question that naturally arises from the analytical expression is: how important is the contribution of the sines? Can we recover the full SFF from just summing over the cosine part, i.e. with a Gaussian approximation for the $k$nLS distribution? The answer is that doing so we get a correlation hole but no linear ramp. So the sines contributions are especially important at the beginning of the ramp and at the transition to the plateau.

The total averaged SFF for the Poissonian ensemble is given exactly by
\begin{eqnarray} \label{SFF_Pois}
    S_t\ep{\text{Poisson}} = \frac{1}{N} +\sum_{k=1}^{N-1}\frac{2(N-k)}{N^2} \frac{\cos(k \arctan t)}{(1+t^2)^{k/2}} .
\end{eqnarray}
Although for $k>1$ the $k$th neighbor SFF for the Poissonian ensemble has a dip and shows some oscillations before flattening out (as can be seen in Figure \ref{fig:SFFk}), the full Poissonian SFF has no `dip' or `correlation hole', as expected for completely uncorrelated levels. It is natural to ask the question: Why does this function not build a correlation, is it because of the $\arctan t$ in the argument of the cosine, or because the envelope is a Lorentzian instead of a Gaussian? It is none of these two reasons. Instead, the main reason is that the width of the Lorentzian decreases as $k$ grows---due to the power $k/2$; similarly, the ramp is absent when the envelope is substituted for a Gaussian with decreasing width, e.g.  $e^{-k t^2}$ or  $e^{-k^2 t^2}$. This ensures that the contribution from large $k$ neighbors is fastly attenuated, effectively restricting their ability to build a ramp. 

\begin{figure}
    \centering
    \includegraphics[width =\linewidth]{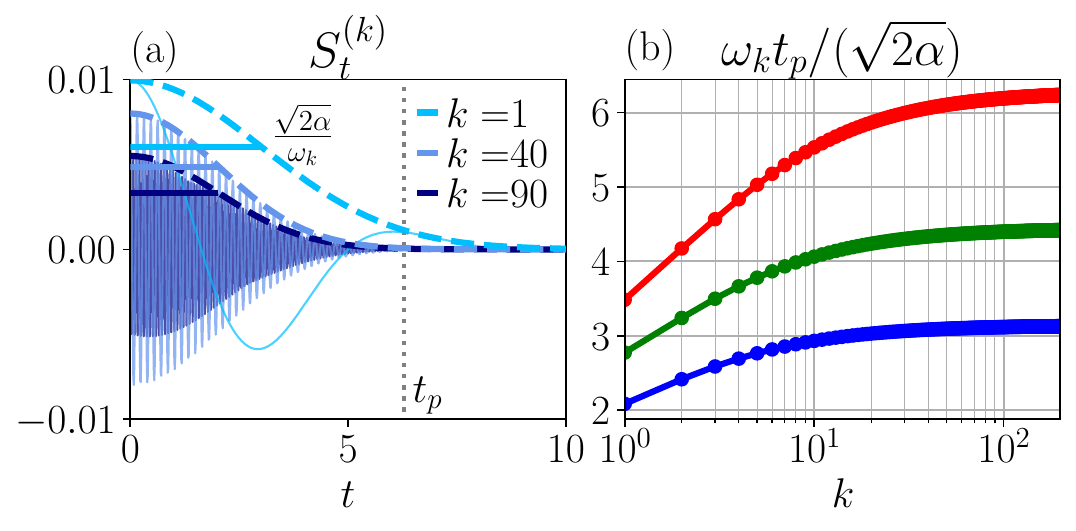}
    \caption{(a) $k$nSFF for GSE for $k=1, \; 40, \; 90$ (solid thin lines) and Gaussian envelope (dashed thick line), standard deviation of the Gaussian envelope $\sqrt{2 \alpha}/\omega_k$ (solid horizontal lines) and plateau time $t_p = 2 \pi$ (dotted gray line). (b) Ratio between the plateau time $t_p=2 \pi$ and the width of the Gaussian envelope $\sqrt{2 \alpha}/\omega_k$ as a function of the neighbor degree $k$ for the three Gaussian ensembles: GOE (red), GUE (green) and GSE (blue). 
    }
    \label{fig:tp_sigma}
\end{figure}

This behavior sharply contrasts with the one observed for RMT. In such case, the width of the Gaussian envelope initially decays but saturates to a constant $\sqrt{2\alpha}/\omega_k \sim \sqrt{\beta}$ for large $k$ as can be observed in Fig. \ref{fig:tp_sigma}(a) for GSE and in (b) for all the ensembles. The slowing down of the Gaussian envelope width ensures that large $k$ neighbors have many oscillations in their $k$nSFF, resulting in the build-up of the ramp.

Note that the fact that the GSE exhibits a spike is related to the Gaussian attenuation that multiplies the sum of cosines and sines in \eqref{SFF_approx}, which has the largest width of the Gaussian ensembles. Indeed, the width is set by $\sqrt{2 \alpha}/\omega_k$, which is proportional to $\sqrt{\beta}$, and the GSE has the largest $\beta=4$. Fig. \ref{fig:tp_sigma} shows the ratio of the plateau time $t_p=2\pi$ to the width of the Gaussian envelope for GOE, GUE and GSE. The envelope width is the largest for the GSE, in which the plateau time lies between 2 and 3 standard deviations of the Gaussian envelope.

\begin{figure*}
	\centering
	\includegraphics[width = \linewidth]{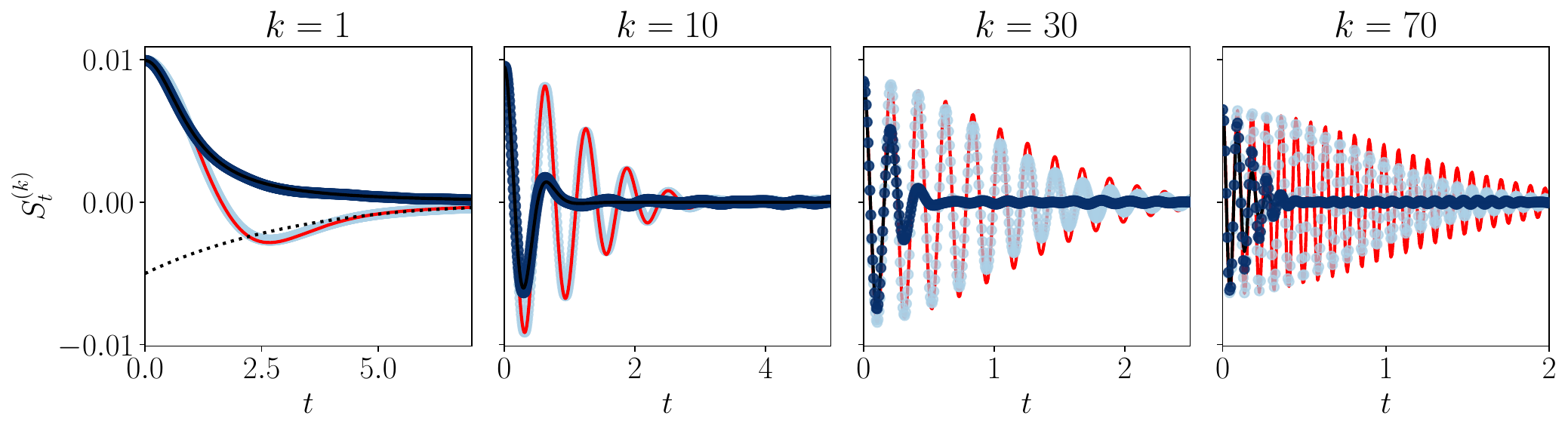}
	\caption{\textbf{$k$-th neighbor Spectral Form Factor for the disordered XXZ spin chain} for different neighbor levels $k=1, 10, 30, 70$ in the chaotic ($W=1$, light blue dots) and the integrable ($W=20$, dark blue dots) phases along with the Poissonian (black line) and GOE (red line) curves. The deviation between the integrable phase and the Poissonian results is apparent starting from $k=10$, and increases for larger $k$. Note that the oscillations differ only in their amplitude but not in their frequency. We emphasis the different scales in the time axis.}
	\label{fig:kSFF_XXZ}
\end{figure*}

\section{Illustration: the disordered XXZ chain\label{sec:XXZ}}
The tools and methodology we introduce in this work are illustrated in a physical model. Specifically, we quantify how closely a many-body quantum system follows the random matrix or Poissonian predictions. 

\subsection{The XXZ spin chain with disorder}

We choose the XXZ spin chain with a varying amount of disorder in the on-site magnetic field, $\hat H =  \hat H_\textsc{xxz} + \hat H_{\text{dis}}$, because this model is known to interpolate between chaos and integrability as a function of disorder strength. The Heisenberg XXZ spin-chain Hamiltonian for $L$ spins reads
\begin{eqnarray}
    \hat H_\textsc{xxz} = \sum_{n=1}^L \big(\hat S_n^x \hat S_{n+1}^{x}+\hat S_n^y \hat S_{n+1}^{y} +J_z \hat S_n^z \hat S_{n+1}^{z} \big),
\end{eqnarray}
where $\hat S_n^{x,y,z}$ are spin 1/2 operators on site $n$. We assume periodic boundary conditions.  This model is known to be integrable and can be solved using the Bethe ansatz \cite{bethe1931theorie}. 
Adding on-site magnetic fields with random strengths 
\begin{eqnarray} \label{H_dis}
    \hat H_{\text{dis}} = \sum_{n=1}^L h_n^z \hat S_n^z,
\end{eqnarray}
where $h_n^z$ are real random numbers taken from a uniform distribution $\mathcal{U}_{[-W/2,W/2]}$, changes the integrability properties of the XXZ chain as a function of the disorder strength $W$. Roughly speaking, when $W$ is small (but not too small), integrability is broken, while as $W$ increases, integrability is restored. In the chaotic regime, the spectral statistics agree with those of GOE due to the system's time-reversal symmetry\footnote{For $\ha{\mc T} = \ha{\mc U} \ha{\mc K}$ the anti-unitary time-reversal operator, generally written as the product of a unitary and complex-conjugation but which can be chosen as $\ha{\mc T} = \ha{\mc K}$ \cite{kawabata_symmetry_2022}, we verify that $\ha{\mc T} \ha H \ha{\mc T}^{-1} = \ha H$ since the Hamiltonian is real. }.
The transition from chaos to integrability in this and similar models can be probed using the SFF \cite{schiulaz_self-averaging_2020, suntajs_quantum_2020}.
The XXZ Hamiltonian with the disorder term \eqref{H_dis} conserves the total spin in the $z$-direction; in other words, it commutes with the operator $\hat{\mathcal{S}}^z = \sum_{n=1}^L \hat{S}_n^z$. 
The Hamiltonian thus does not mix sectors of different $\hat{\mathcal{S}}^z$ eigenvalues, and we can work in one such sector.  We choose to work in the sector with half of the spins up and half of the spins down, which is of dimension ${{L}\choose{L/2}}$.  We present results for $L=16$ for which the Hilbert space dimension in the above-mentioned sector is $12{,}870$.
In practice, however, we draw our statistics from  $N=200$ eigenvalues around the densest part of the spectrum.  

\subsection{Dynamical signatures of chaos ($k$nSFF) in the disordered XXZ}
To test how the dynamical signatures of a real system match those predicted by RMT or Poisson, we extract data for the $k$nSFF from the XXZ model with disorder. 
Figure \ref{fig:kSFF_XXZ} shows numerical results for the $k$th neighbor SFF for various values of $k$, where  the behavior for a disorder strength of $W=1$ can be compared with  $W=20$. \pablo{The plot compares the spin chain numerical data with those from the Poisson and GOE ensembles. In particular, we see that at small $k$ the XXZ dynamics agrees very well with the analytical result. However, we observe a more pronounced deviation from both GOE and Poisson at large, i.e. $k = 30$ and specially $k=70$. 
The period of the oscillations is well captured by the analytical expressions, but the deviations from the analytical $k$nSFFs show up in the envelope, i.e. in the amplitude of the oscillations, which is narrower than GOE for $W=1$ and broader than Poisson for $W=20$.} This is a well-known behavior for other long-range measures of spectral correlations, which stop following the universal predictions of Random Matrices and Poisson.

\begin{figure}
    \centering
    \includegraphics[width = \linewidth]{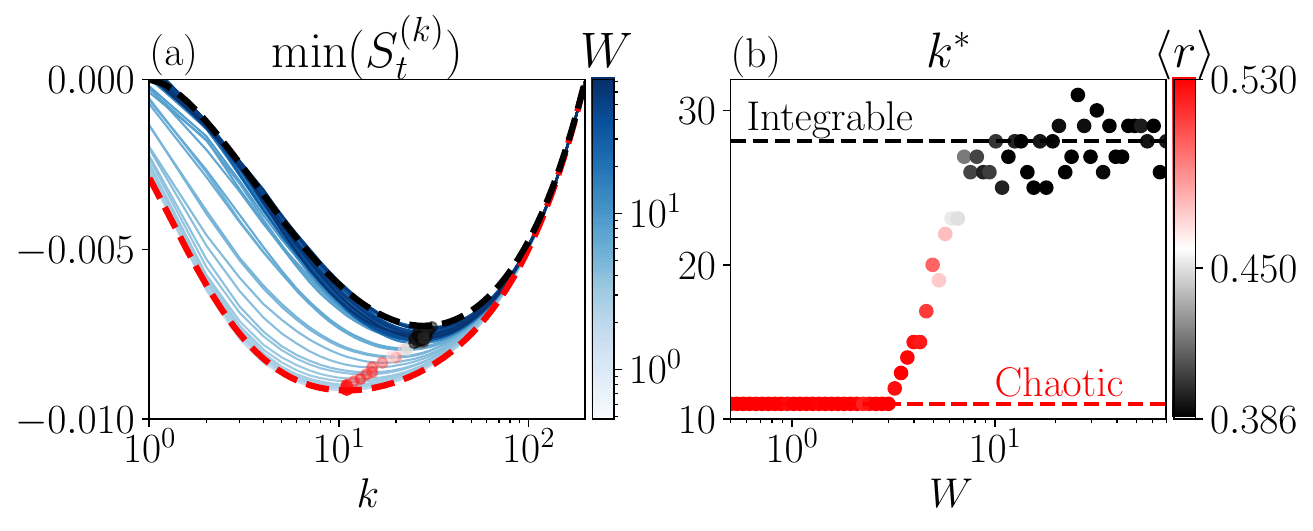}
    \caption{(a) \textbf{Minimum of the $k$nSFF as a function of the neighbor degree $k$}, the colored dots mark the deepest $k$nSFF, (b) \textbf{deepest $k$-th neighbors SFF $k^*$ as a function of the disorder strength}, the colorscale marks the $\av{r}$ parameter \cite{atas2013distribution}. The values of $k^*$ for GOE and Poisson are $k^* = 11$ and $k^* = 28$ respectively, which can be obtained from the expansions for $k^*$ \eqref{eq:kstar}, \eqref{eq:kstar_poi} respectively with $N=200$ which is the energy window size. }
    \label{fig:XXZ_Stdip}
\end{figure}

Figure \ref{fig:XXZ_Stdip}(a) shows the results for the minimum value $S\ep{k}_{t_m}$ as a function of $k$ as well as the values of $k^*$ as a function of the disorder strength $W$. The minimum of the $k$nSFF agrees very well with the Poisson and GOE limits because, as discussed previously and seen in Fig.~\ref{fig:kSFF_XXZ}, the most relevant corrections to the knSFF are in the envelope, and thus are only for longer times, comparable to the width of the envelope. Furthermore, for $k=10$ the chaotic phase of the XXZ model follows well the GOE prediction and for $k=30$ the integrable phase of XXZ agrees with the Poisson prediction, as seen in Fig. \ref{fig:kSFF_XXZ}. Deviations from the limiting behavior will not be apparent there.
Therefore, the minimum attained by the $k$nSFF is a quantity which, even if it can give information about the transition between chaos and integrability, is not sensitive to deviations of universality in long-range correlations $k \gg 1$.

The deepest neighbor $k^*$ shows different behavior in the chaotic and integrable phase, and thus can be used to probe the transition, in a similar way to the level spacing ratios $\langle r \rangle$ \cite{atas2013distribution}. This is shown in Fig. \ref{fig:XXZ_Stdip} (b) where as we increase $W$ we go from a value of $k^*=11$ to a value of $k^*=28$.  These values of $k^*$  are exactly those predicted from the GOE and Poissonian ensembles and respectively correspond to chaotic (small $W$) and integrable (large $W$) dynamics.

\begin{figure}
    \centering
    \includegraphics[width=0.6\linewidth]{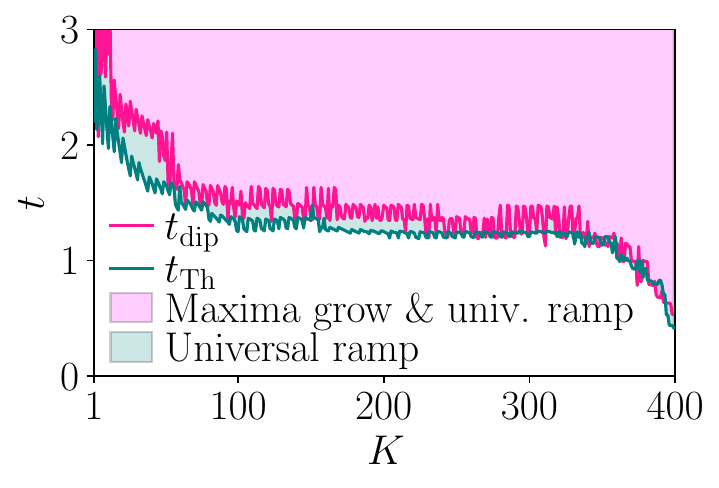}
    \caption{\textbf{Thouless and dip times} for the $k$nSFF as a function of the maximum neighbor range $K$ for the XXZ model in the chaotic phase. The results use $\epsilon = 0.2$ \cite{suntajs_quantum_2020} considering $N_\mr{en}=400$ energy levels and averaged over $N_\mr{av}=150$ realizations of the disorder. }
    \label{fig:XXZ_tho_dip}
\end{figure}

Figure \ref{fig:XXZ_tho_dip} shows the dip and Thouless times of the partial SFF for the XXZ spin chain with disorder in the chaotic phase. Similarly to the behavior observed in random matrices we see that both of these times decrease as the maximum number of neighbors $K$ is increased. Another interesting behavior, also shown by RMT, is that the short $K \lesssim 100$ and the very long range $K \gtrsim 300 $ neighbors are the ones that contribute the most to the length of the ramp of the SFF, indeed Fig. \ref{fig:XXZ_tho_dip} shows that both the Thouless and dip times stay almost constant as we increase the maximum range of neighbors considered from $K \approx 100$ to $K \approx 300$. Counterintuitively, even if these long ranges do not show RMT universality, they are key to explain the full extent of the ramp observed in the SFF. This suggests that either the ramp involves some contributions from long-range neighbors which do not show much difference with RMT, or that the deviations from universality cancel out to give the universal ramp.  Interestingly, for the disordered XXZ we do not see that the dip time precedes the Thouless time, but remarkably the two curves show much more similarity that for GOE. 
 
\section{Discussion} \label{sec:discussion}

In this work, we studied the role played by the short, medium, and long-range spectral correlations in building the SFF. Introducing the $k$-th neighbor SFF, we found analytical results for it in the three Gaussian ensembles (GOE, GUE, and GSE) and the Poissonian ensemble. 
It is conjectured that  realistic many-body systems, as illustrated by the disordered XXZ spin chain, will show similar spectral statistics \cite{bohigas_characterization_1984} as these ensembles (GOE and Poisson) in their chaotic and integrable phases, and interpolate between them in the transition. This correspondance breaks down when long-range correlations are considered, as measured e.g. by the number variance \cite{bertrand2016anomalous}, which are known to show deviations from universality. Our results provide an alternative point of view to study the breakdown of Random Matrix universality for long-range correlations, focusing on the role that non-universal correlations play in building the universal ramp of the SFF.

Leveraging the $k$-th neighbor level spacing surmise, we found expressions for the $k$-th neighbor SFFs for the random matrix ensembles and for the Poissonian ensemble. The Gaussian ensembles $k$nSFF's can be approximated by a sum of cosines and sines with appropriate polynomial coefficients and an overall Gaussian envelope function. We first characterized the properties of single $k$nSFFs, providing expressions for: its minimum value, the time at which the minimum happens, and the energy range for which $k^*$nSFF is the deepest. In general, these quantities depend on the dynamical phase of the system, whether it is chaotic or integrable. However, the minimum time $t_m(k)$ does not distinguish between chaos and integrability for $k\geq 2$. The minimum value attained by the $k$nSFF's is smaller in the random matrix ensembles than for Poisson, and the deepest $k$nSFF shows a different scaling $k^* \sim N^{1/3}$ for random matrices and $k^* \sim N^{1/2}$ for Poisson. These three features of the $k$nSFF's are not sensitive to deviations from RMT universality since the minimum time (cf. Fig. \ref{fig:comparison_num_rmt} (b)) and the minimum value and deepest $k^*$ (cf. Fig. \ref{fig:XXZ_Stdip}), computed for the disordered XXZ along the chaos to integrability transition, accurately follow the RMT results. This can be understood since the single $k$nSFFs show deviations from universality in the width of the envelope, which are very small at short times. 


For the Gaussian ensembles, the several approximations we made to achieve simple, tractable expressions capture the main properties of the complete SFF. The role of each $k$-th neighbor level spacing in the full SFF (cf. Sec. \ref{sec:building_SFF_fromknSFF}) is investigated using the partial SFF considering only spectral neighbors in the range $[1, \dots, K]$ and in particular introducing two time scales that determine the onset of the ramp, the \textit{dip} and \textit{Thouless} times. For the different ensembles of Random Matrices, we observe that these times decrease as more and more neighbors are accounted for. However, they show a particularly fast growth for the longest range spacings, when $K$ is close to the number of energy levels considered. A similar behavior is also seen for XXZ in the chaotic phase, where the biggest decrease in the characteristic time-scales of the onset of the ramp happens for short range level spacings and for very long range level spacings, with an almost constant onset of the ramp from $K \approx 100$ to $K \approx 300$. Interestingly, the dip and Thouless times of the partial SFF are much more similar for the disordered XXZ model than for RMT. Surprisingly, when only even or odd neighbor ranges are considered, the SFFs change drastically, with even neighbors building most of the correlation hole and having a constructive interference at $t=\pi$, while the odd neighbors only contribute to cancel the constructive interference. Lastly, we also studied how the SFF computed from the approximated $k$nSFF's builds the full SFF. We compared it with numerical results and the analytics given by the connected part. In particular, the Gaussian envelope times oscillating function shape of the $k$nSFF's allows to understand why GSE shows the spike right before the plateau, because the envelope has not decayed enough to avoid all the $k$nSFF's constructively interfering to give the spike at $t_p = 2\pi$.


Possible future research directions include the study of how finite temperature affects the $k$nSFFs \cite{martinez-azcona_analyticity_2022}, the extension of the $k$nSFFs to the dissipative case, and for which only nearest-neighbor correlations and SFF have mostly been studied \cite{sa_complex_2020, xu_extreme_2019}. Lastly, in light of our results a natural question arises: which functional forms of the $k$nSFF's build a linear ramp with a correlation hole and which do not? We have so far only argued that the narrow Lorentzian envelope in the Poissonian ensemble was hindering any possible buildup of a correlation hole. However it remains to be understood why the Gaussian RMT expressions build a \textit{linear ramp}.
We also leave to future research the investigation of the $k$th neighbor autocorrelation functions defined in this work. In particular, they can be used to better understand the behavior of the autocorrelation function for different operators. 

To summarize, our results suggest that correlations beyond first energy neighbors have a clear manifestation in dynamical quantities such as the spectral form factor, and are key to explain the full extent of the ramp. The specific nature of the level repulsion beyond just nearest-neighbor eigenenergies plays an important role in accurately capturing the complex features of many-body quantum systems at all time scales.

\begin{acknowledgments}
We would like to thank Lea F. Santos, Federico Balducci, Andr\'as Grabarits, Aritra Kundu, Oskar Prosniak, and Federico Roccati for insightful discussions and comments on the manuscript.  Some of the numerical simulations presented in this work were carried out using the HPC facilities of the University of Luxembourg. This work was partially funded by the John Templeton Foundation (Grant 62171)  and the Luxembourg National Research Fund (FNR, Attract grant 15382998). The opinions expressed in this publication are those of the authors and do not necessarily reflect the views of the John Templeton Foundation.
\end{acknowledgments}






\appendix
\section{Details on the dip and Thouless times of the partial SFF}
\label{app:dip_time_SFF}
\begin{figure}[h]
    \centering
    \includegraphics[width=.75\linewidth]{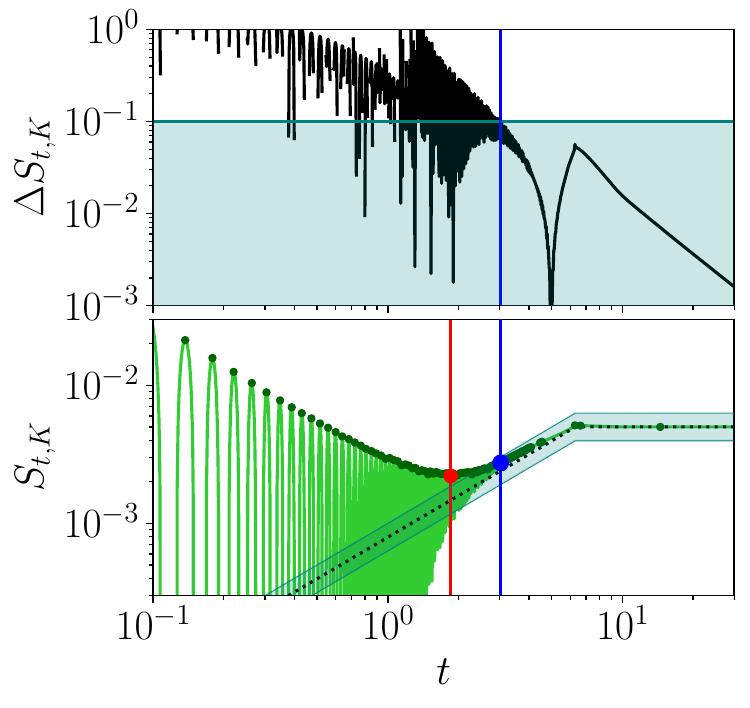}
    \caption{Visualization of the Thouless (blue) and dip (red) times for the partial SFF with $K= 150$, $N=200$ for the GUE, constructed from the analytical expressions for the knSFF's derived in the main text. (a) Visualization of the logarithm of the ratio of SFF and the connected SFF $\Delta S_{t, K}$. The tolerance parameter $\epsilon=0.1$ (turquoise horizontal line) and the Thouless time (blue vertical line). (b) Partial SFF for the GUE (light green solid line), along with the relative maxima (green circles), and the connected SFF (black dotted line). The shaded area shows the interval $[b_t 10^{-\epsilon}, b_t 10^{+\epsilon}]$ for visualization of the condition $\Delta S_{t, K} = \epsilon$. We stress out that the defining condition of the Thouless time is more clearly seen in the behavior of $\Delta S_{t, K}$ and not in this area.}
    \label{fig:thoulesDip_t_vis}
\end{figure}

In this Appendix, we give further details on the Thouless and dip times of the partial SFF. Figure \ref{fig:thoulesDip_t_vis} (a) shows $\Delta S_t\ep K$ as defined in \eqref{eq:def_DeltaStk} for the partial SFF with $K=150$ neighbors for the GUE computed from the approximation of the knSFF. The conditions to define the Thouless time ensure that the ratio between the partial SFF $S_{t, K}$ and the connected SFF $b_\ts{gue}(t)$ remain close, within a window defined by the tolerance parameter $\epsilon$. This plot also shows that the distance between the sum of some of the approximated knSFFs and the connected SFF remains small, but has a minimum before the onset of the plateau at $2 \pi$, a maximum exactly at the Heisenberg time $t_p = 2 \pi$, and decreases in the plateau. This suggests that higher neighbors, with $k>150$ in this case, play a role in the transition from the ramp to the plateau of the SFF. 

\section{The connected SFF for Gaussian ensembles and sine-kernel} \label{App:SFF_connected}

The 2-point spectral connected correlation function $\rho_c(E, E')$ of a certain ensemble is defined as \cite{haake_quantum_1991}
\begin{equation}
    \rho_c(E, E') = \Av{\rho(E) \rho(E')} - \av{\rho(E)}\av{\rho(E')},
\end{equation}
where $\rho(E) = \sum_j \delta(E - E_j)$ and $\av{\bullet}$ represents a suitable average, e.g. over the random matrix ensemble. Thus $\av{\rho(E)}$ is the average density of states. After unfolding, i.e. introducing the rescaled energies $\varepsilon = E N \av{\rho(E)}$, the renormalized 2-point correlation function reads
\begin{equation}
    \frac{\rho_c(E, E')}{\av{\rho(E)}\av{\rho(E')}} = \delta(\varepsilon - \varepsilon') - Y(\varepsilon, \varepsilon'),
\end{equation}
where $Y(\varepsilon, \varepsilon')$ is \textit{Dyson's two-level cluster function}, defined as
\begin{equation}
    Y(\varepsilon, \varepsilon') = 1 - \Av{\sum_{m \neq n} \delta (\varepsilon - \varepsilon_n) \delta(\varepsilon'-\varepsilon_m)}. 
\end{equation}
This function depends only on the difference $s = \varepsilon - \varepsilon'$ and for the GUE in the large $N$ limit reads \cite{haake_quantum_1991}
\begin{equation}
    Y^\ts{gue}(s) = \mr{sinc}^2(\pi s),
\end{equation}
where the sinc function is defined as $\mr{sinc}(x) = \sin(x)/x$. This is typically known in the literature as the sine-kernel. The 2-point correlation function is closely related to it through $R_2(s) = 1 - Y_{2}(s)$, which can be obtained as a sum of the knLS level spacing distributions \cite{pandey2019quantum}
\begin{equation}
    R_2(s) = \sum_{k=1}^{\infty} \mc P\ep k (s).
\end{equation}


The connected SFF is defined as the Fourier transform of the connected correlation function. For the random matrix ensembles it can be obtained from the cluster function, for GUE  it reads \cite{haake_quantum_1991}
\begin{equation}
    b_\ts{gue}(t) = \begin{cases}
         \frac{t}{2 \pi N} & \mr{for}\; t \leq 2 \pi \\ 
         \frac{1}{N} &  \mr{for}\; t > 2 \pi,
    \end{cases}
\end{equation}
where we adapted it to our Heisenberg time of $t_p=2 \pi$ and the plateau value of $\lim_{t \rightarrow \infty} S_t = 1/N$.
For GOE it reads
\begin{equation}
    b_\ts{goe}(t)=  \left\{ \begin{array}{ll}
          \frac{t}{\pi N} - \frac{t}{2 \pi N} \log(1 + \frac{t}{\pi}) & \mr{for}\; t \leq 2 \pi \\[\smallskipamount]
         \frac{2}{N} - \frac{t}{2 \pi N}\log \frac{t + \pi}{t - \pi} &  \mr{for}\; t > 2 \pi.
    \end{array}\right. 
\end{equation}

And lastly, for the GSE, it is 
\begin{equation}
    b_\ts{gse}(t)=  \left\{ \begin{array}{ll}
          \frac{t}{4 \pi N} - \frac{t}{8 \pi N} \log(1 - \frac{|t|}{2\pi}) & \mr{for}\; t \leq 4 \pi \\ [\smallskipamount]
         \frac{1}{N} &  \mr{for}\; t > 4 \pi.
    \end{array}\right.
\end{equation}


\section{Unfolding the spectrum} \label{App:unfolding}
The spectrum of a system is a property which \textit{a priori} depends on the system under consideration. However, its energy correlations can obey some universal laws. To study the latter, we need to remove the dependence of the spectrum on non-universal features, like the density of states $\bar \rho (E)$.
In doing so, systems that are originally completely different can be compared. 
The procedure to remove the dependence on the local density of states is known as \textit{unfolding} (see e.g. \cite{Guhr:1997ve}). In this appendix, we explain how we unfold a generic spectrum. We then study some aspects of the effect of unfolding on our results. 

Our method of unfolding involves computing the function
\begin{eqnarray} \label{fE}
    f(E) = N \int_{-\infty}^E dE' \, \bar{\rho}(E')~,
\end{eqnarray}
where $\bar{\rho}(E)$ is the average density of states, and then pass the energy eigenvalues $\{E_i\}_{i=1}^N$ into this function to get the set of \textit{unfolded} energy levels $\{e_i\}_{i=1}^N = \{f(E_i)\}_{i=1}^N$. 
For the random matrix ensembles we study, the average density of states is given by the Wigner semicircle distribution, $\bar{\rho}(E) = \frac{1}{\pi \beta N} \sqrt{2 N \beta - E^2}$. This leads to an analytical form for the function $f(E)$, which reads \footnote{This result can be found in \url{https://robertsweeneyblanco.github.io/Computational_Random_Matrix_Theory/Eigenvalues/Wigner_Surmise.html}}
\begin{equation}
    f(E) = \frac{N}{2}+ \frac{1}{\pi \beta} \left( N \beta \arcsin \ \frac{E}{\sqrt{2 N \beta}} + \frac{E}{2}\sqrt{2 N \beta - E^2}\right),
\end{equation}
for $- \sqrt{2 \beta N} < E < \sqrt{2 \beta N}$, while it is $f(E) = 0$ for $E\leq - \sqrt{2 \beta N}$ and $f(E) = N$ for $E \geq \sqrt{2 \beta N}$, see also \cite{abul-magd_unfolding_2014}.

For the disordered XXZ spin chain, there is no analytical expression for the average density of states. We thus rely on a \textit{numerical} polynomial fit for $f(E)$ for each realization of disorder. We used a larger window of energies to perform the fit, and then discarded the two edges, thus focusing our analysis on a window of around $N=200$ energies. 
\begin{figure}[h]
    \centering
    \includegraphics[width = .8 \linewidth]{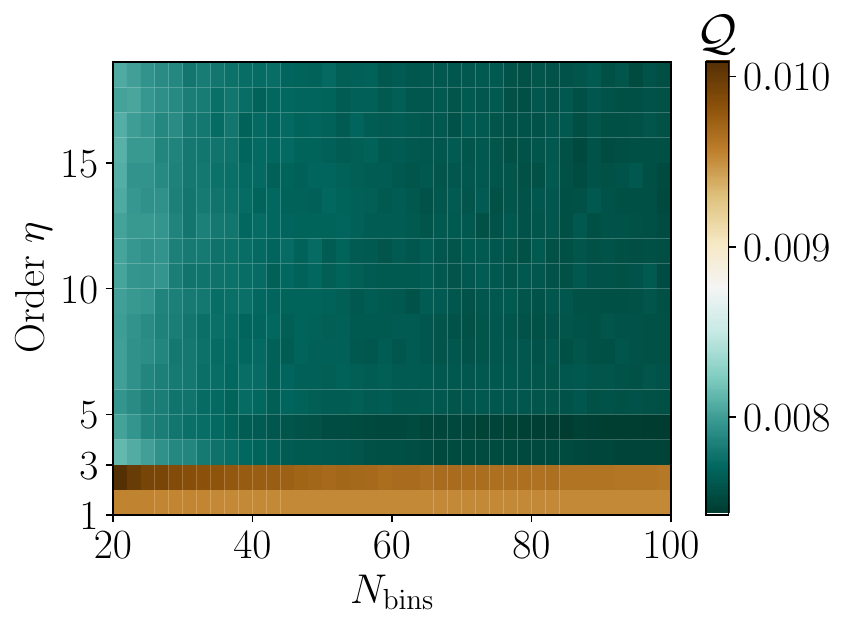}
    \caption{Quality of the unfolding as a function of the order and the number of bins for the GOE.}
    \label{fig:rmt_fit_unf}
\end{figure}

Our numerical unfolding depends on two parameters: the maximum order of the polynomial fit $\eta$ and the number of bins with which we construct our histogram (related to the bin's width). These parameters, especially the polynomial order, can critically change the results since we can be overfitting the spectrum and include some of the universal correlations into the density of states. To check which minimum order gives a reasonable fit, we compare the numerical and analytical unfolding on a random matrix and define a \textit{quality of the fit}, $\mc Q$, as the square of the difference between the histograms of the analytical and numerical unfolded spectra, namely
\begin{equation}
    \mc Q = \sum_{n \in \mr{bins}} \left[\mr{Hist}_n(f^\mr{ana}(E))- \mr{Hist}_n(f^\mr{num}(E)) \right]^2.
\end{equation}
Figure \ref{fig:rmt_fit_unf} shows that $\eta = 3$ is an unfolding order with already good results. So, we chose this order to avoid over-fitting. The parameter of the number of bins is not too critical, and we set $N_\mr{bins}=50$ to have enough bins and enough points per bin.

\begin{figure*}
    \centering
    \includegraphics[width = .75\linewidth]{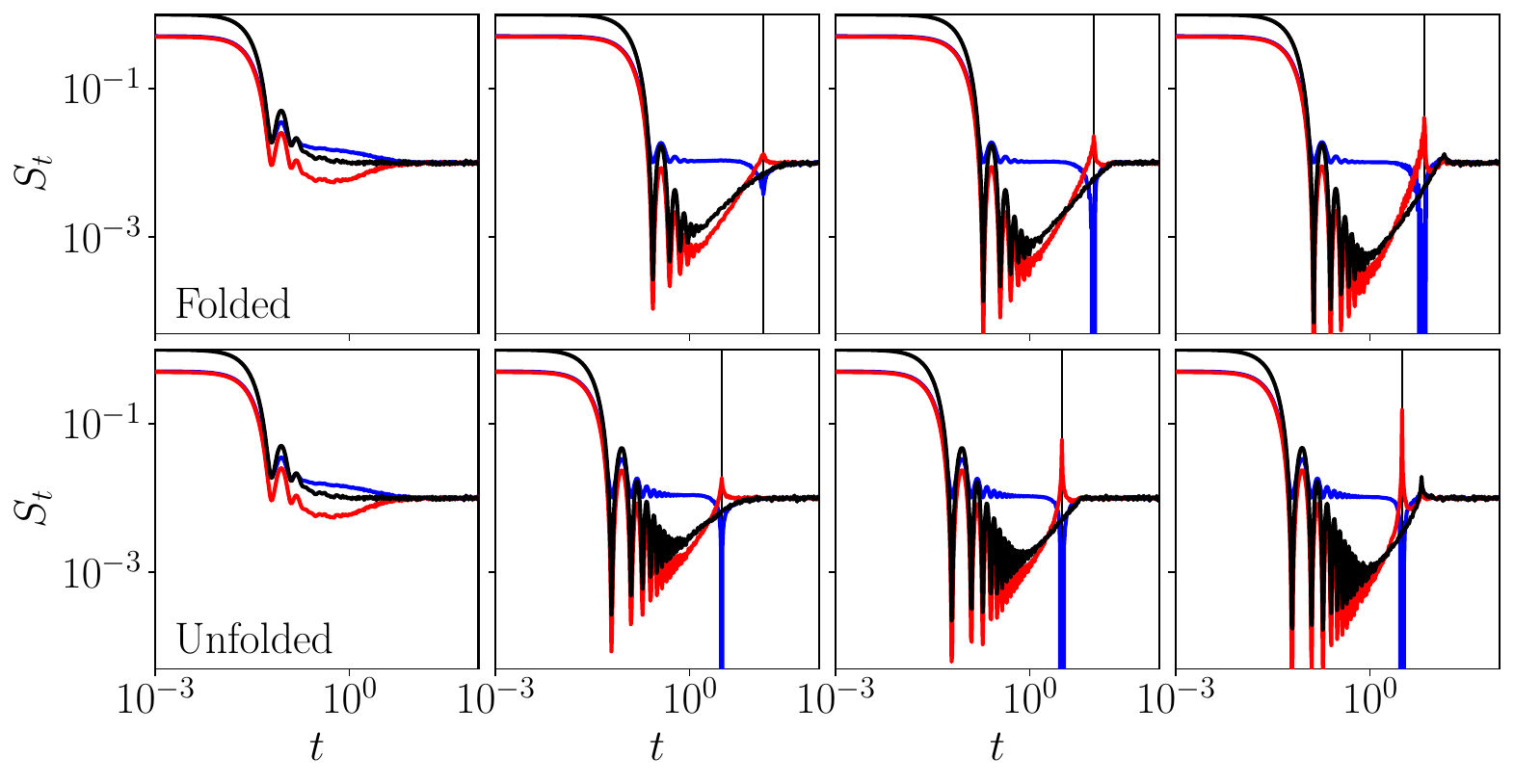}
    \caption{The even- (red) and odd- (blue) neighbor contributions are compared for the folded and unfolded cases of Poisson, GOE, GUE, and GSE (left to right). The black solid line is the total SFF. The figure shows numerical data from random matrices of dimension $N= 100$ averaged over $N_{\textrm{av}}=1000$ realizations. }
    \label{fig:EvenOdd_FoldvsUnFold}
\end{figure*}

Although our results refer to the unfolded spectrum, we study here some of the same results for the folded (i.e. the original, \textit{not} unfolded spectrum), namely, the even-odd signatures discussed in Section \ref{sec:even_odd}.  
Figure \ref{fig:EvenOdd_FoldvsUnFold} confirms that such a signature is still present in Random Matrices without unfolded spectrum. The main difference is in the time at which the `resonance' and `anti-resonance' appear: while for the unfolded spectrum, they appear at $t=\pi$, for the folded spectrum, they appear at a time scale related to the matrix dimension.

\section{Self-averaging of the $k$th neighbor SFF} \label{app:self-av}
In this appendix, we discuss the self-averaging properties of the $k$th neighbor SFFs, as a function of $k$.

A quantity is said to be self averaging if its relative variance becomes smaller as the system size is increased. 
The SFF is known to be particularly not self-averaging around its plateau (the flat part which the SFF tends to at large $t$), i.e. the relative variance of its plateau increases as $N$ is increased.  
Here, we numerically study the self-averaging of the plateau of the $k$th neighbor SFF. The relative variance of the $k$th neighbor SFF can be defined as (see e.g. \cite{schiulaz_self-averaging_2020}):
\begin{equation}
    R_k(t) = \frac{\av{(\mathcal{S}_t\ep k +\bar{\mc S} )^2}-\av{\mathcal{S}_t \ep k+\bar{\mc S} }^2}{\av{\mathcal{S}_t \ep k+\bar{\mc S} }^2} ~,
\end{equation}
where $\bar{\mc S} = \tfrac{1}{N(N-1)}$ is the value of the plateau divided equally among the $N-1$ possible neighbors, which we added such that the average $\av{\mathcal{S}_t \ep k+\bar{\mc S} }$ is  non-zero.

\begin{figure}
    \centering
    \includegraphics[width = 0.85\linewidth]{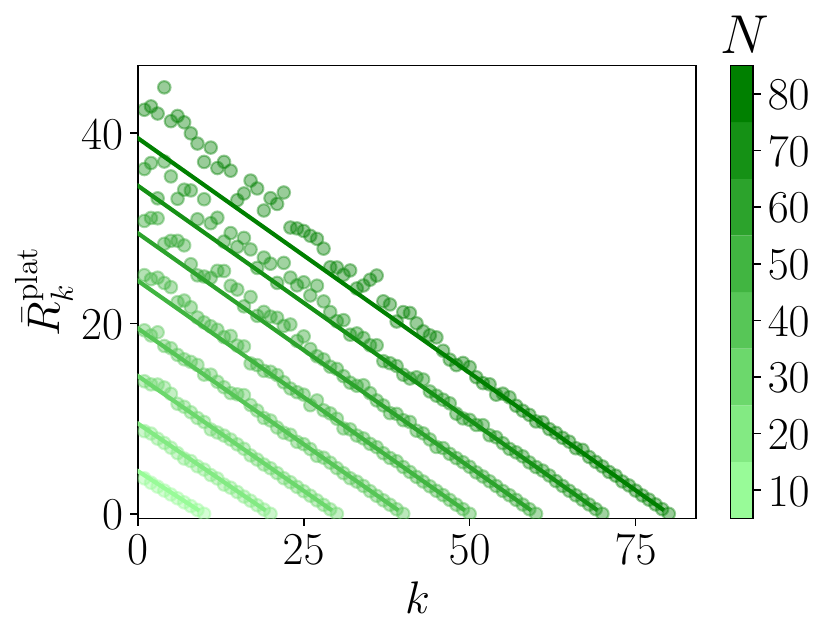}
    \caption{Relative variance of the plateau of the $k$nSFF as a function of the neighbor degree $k$, for GUE random matrices of dimension $N$ (colorscale). The results are averaged over $N_\mathrm{av}=1000$ ensemble realizations and $\bar R^\mathrm{plat}_k:= \frac 1 T \int_{t_p}^{t_p+T}R_k(\tau) d \tau$  is the time averaged relative variance on the plateau. Solid lines correspond to \eqref{eq_RV_plat}.}
    \label{fig:relVar_plat}
\end{figure}

Figure \ref{fig:relVar_plat} shows the relative variance of the plateau as a function of the neighbor degree $k$ for different dimensions of the random matrices. We can observe several features:
\begin{itemize}
    \item $\mathcal{S}\ep k _t$ is never self-averaging since the relative variance increases with the dimension of the matrix.
    \item The limiting value of the relative variance decreases linearly with $k$ in the following way
    \begin{equation}\label{eq_RV_plat}
        \bar R_k^\mr{plat} = \frac{1}{T}\int_{t_p}^{t_p+T}R_k(\tau) d \tau =  (N- k)\frac{N-1}{2 N},
    \end{equation}
    where indeed Figure \ref{fig:relVar_plat} suggests a linear function of $k$ with a constant slope independent of $N$ and a constant which scales linearly with $N$.
\end{itemize}

\begin{figure*}
    \centering
    \includegraphics[width = \linewidth]{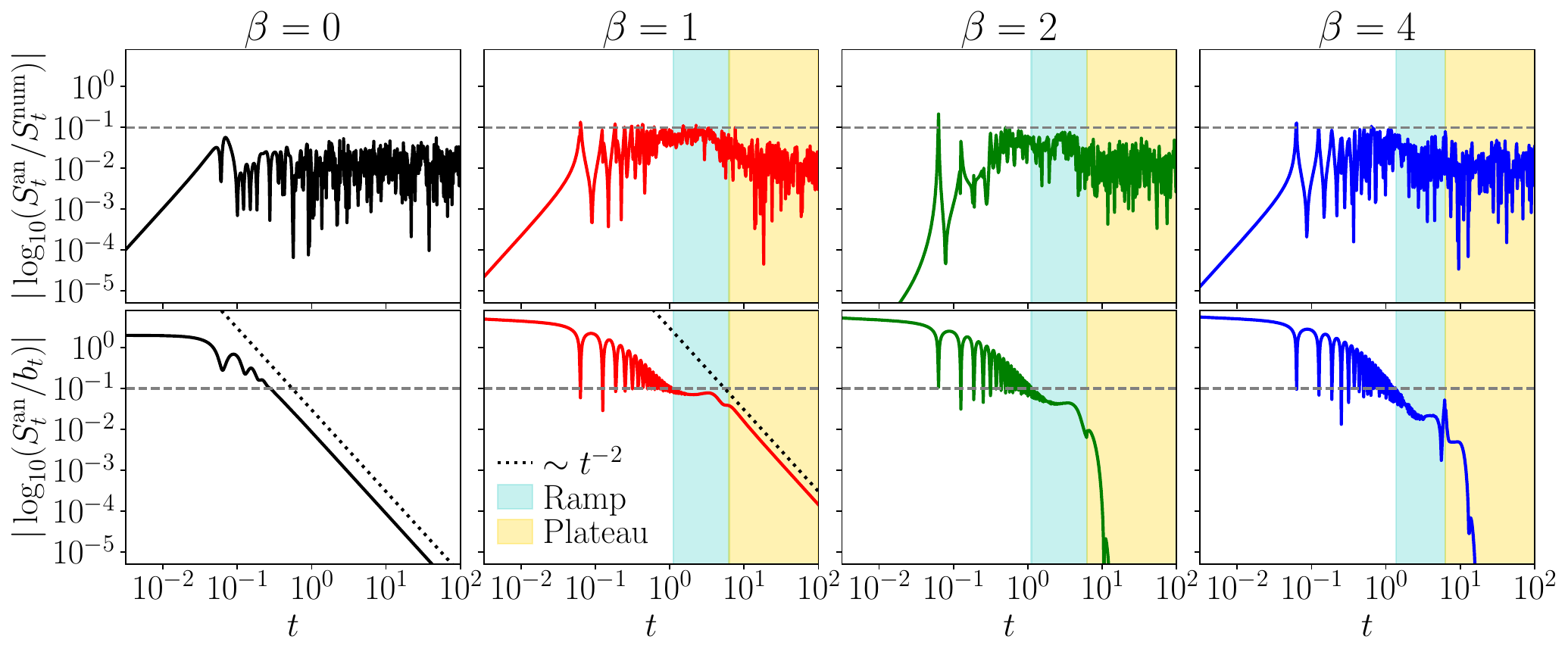}
    \caption{\textbf{Comparisons of the approximate analytical SFF with numerical (top) and connected (bottom) SFF's} for (left to right): Poisson, GOE, GUE and GSE. The dashed line marks the $\epsilon = 0.1$ difference between the two quantities. We observe that the difference with the numerics stays bounded by 0.1, i.e. a $10\%$ difference between the two SFF's. The colored regions indicate the universal ramp (light blue), from the Thouless time to the Ehrenfest time, and the plateau (light yellow), after the Ehrenfest time.  }
    \label{fig:SFF_test_approx}
\end{figure*}

\section{Test of the approximations for the SFF}\label{App:SFF_test}
Since we have made several approximations on the way to our final expressions for the total SFF for the Gaussian ensembles, Eq. \eqref{SFF_approx}, we test the validity of our approximate analytical expression. Figure \ref{fig:SFF_test_approx} (top) shows the relative error between the numerical SFF (computed for RMT) and the expression \eqref{SFF_approx} for the three Gaussian ensembles, as well as a comparison between the Poissonian expression \eqref{SFF_Pois} and numerical data for the SFF taken from the Poissonian ensemble. Following the discussion of the Thouless time \cite{suntajs_quantum_2020}, we decide to take $|\log(f(x)/g(x))|$ as our notion of the distance between the two SFFs. We see that in GOE and GUE the approximation over-estimates slightly the ramp, and especially in GOE has a distance to it of around 10\% of the value of the SFF at that point. However, the distance between the two SFFs always remain bounded by 0.1, indicating that the approximations derived in the main text work reasonably well. 

A further interesting comparison is the SFF obtained from the analytical approximations to the connected SFF, see Fig. \ref{fig:SFF_test_approx} (bottom). This defines the Thouless time for Random Matrices, as the time after which our analytical expression remains close to the universal connected SFF. We observe that at short times, in the initial slope of the SFF the distance between the SFF and the coonnected part is big since at short times $b_t$ is very small. Then we see a transition to the ramp, indicated in blue, where the distance remains below a certain tolerance parameter $\epsilon = 0.1$. The onset of the ramp is defined from the Thouless time \cite{suntajs_quantum_2020}. At $t_p=2 \pi$, we have a transition to the plateau. For GOE we see a power law $\sim t^{-2}$ convergence to the plateau, also seen for Poisson. Interestingly, for GOE and GSE this convergence is much faster than this power law, we also observe that the onset of the plateau is smooth for GOE and becomes less smooth for GUE and GSE. This is due to the fact that recovering the kink in GUE, and especially the spike in GSE, is hard with our approximations to the SFF. However, this distance remains small. The spiky behavior observed for GSE implies that the value of the tolerance parameter $\epsilon$ should be taken carefully so that it characterizes the ramp, and not just the plateau.

\section{Toy model: the complete SFF with only nearest-neighbor correlations}\label{NN_corr}
As we have shown in the main text, the nearest-neighbor level spacing, although indicative of chaotic or regular behavior, is not a sufficient condition for chaos, since truly chaotic models (as modeled by RM) have correlations all over the spectrum. In this spirit, we construct a toy model which only has energy correlations to nearest neighbors, but nowhere else in the spectrum. What would be the SFF of such a system? To answer this, let us recall that the probability distribution of the sum of two uncorrelated random variables $z = x+y$ is given by their convolution. So, in this toy model, the second level spacing distribution simply reads 
\begin{align}
    \mc P\ep{2} (s\ep{2}) &= \mc P\ep{1}(s\ep{1})* \mc P\ep{1}(s\ep{1})\\
    &= \int_0^{s\ep 2} \mr d s \mc P\ep{1}(s) \mc P\ep{1}(s\ep{2}-s). \notag
\end{align}

The convolution theorem states that the Fourier transform of a convolution is the product of the Fourier transform, and vice versa. The $k$nSFF for this toy model follows as
\begin{equation}
    S\ep k_t = \mr{Re}(\mc F[\mc P\ep 1]^k(t)),
\end{equation}
where the Fourier transform of the nnLS distribution, $\mc F[\mc P\ep 1](t)$,  admits the exact expression
\begin{align} \label{SFF_nnCor}
    \mc F[\mc P\ep 1](t) = & \:{}_1F_1\left(\frac{\beta+1}{2}, \frac{1}{2}, -\frac{t^2}{4 A_\beta}\right) \\ \notag &-i t  \; {}_1F_1\left(\frac{\beta}{2}+1, \frac{3}{2}, -\frac{t^2}{4 A_\beta} \right). 
\end{align}
\begin{figure}
    \centering
    \includegraphics[width = .7\linewidth]{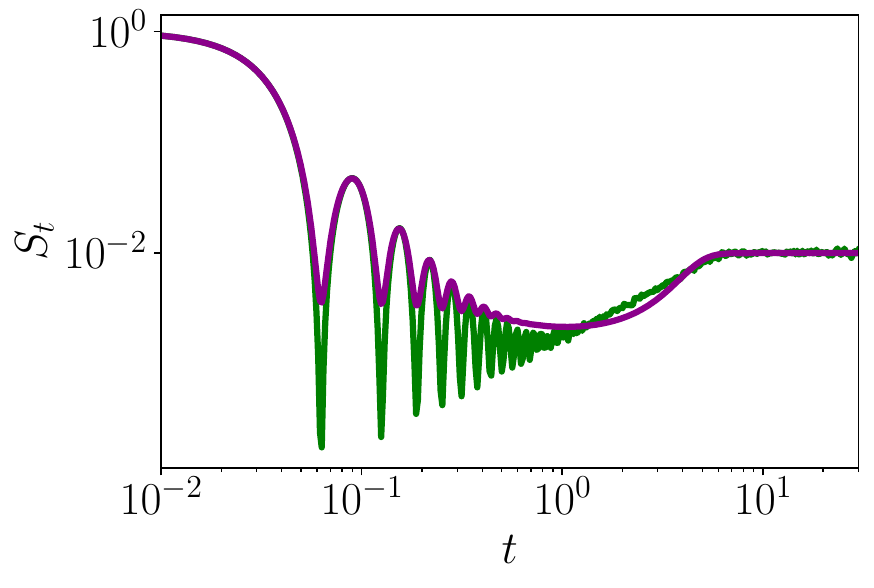}
    \caption{SFF computed numerically for the GUE (green) and for the toy model with energy correlations to nearest neighbors only (purple), Eq. \eqref{SFF_nnCor}. }
    \label{fig:SFF_uncorrel}
\end{figure}

The sum of the $k$nSFF is shown in Fig. \ref{fig:SFF_uncorrel} for the GUE ensemble. The SFF of this toy model shows a correlation hole since it decays and grows back, but the ramp is not linear and therefore there is no chaos. Similar non-linear ramps in the SFF have been reported for integrable models like the SYK$_2$ \cite{winer_exponential_2020, liao_many-body_2020}. Thus, we conclude that correlations beyond the nearest energy levels are needed to find the linear ramp in the SFF characteristic of chaotic systems.

\section{A dissipative protocol to measure the $k$nSFF}
\label{sec:protocol_knSFF}
The autocorrelation function of a general operator $\ha O$  \eqref{eq:correl_fct_def} can be obtained from knowledge of the spectrum and the operator. We propose a protocol, based on dissipative dynamics, to measure the $k$-neighbor autocorrelation function.

Assume that we are able to prepare an initial operator with non-zero weight only in its main and $k$-th diagonal, namely $\ha O\ep k = \sum_{n=1}^N O_{ii}\ket{i}\bra{i} + \sum_{i=1}^{N-k} O_{i+k, i} \ket{i+k}\bra{i} + \textrm{h.c.}$. Its autocorrelation function will be related to the $k$nSFF since it only contains spectral information from the $k$nLS,  
\begin{align} 
    \mc C_t\ep k &= \frac{1}{\mc N^2}\mr{Tr}(\ha O\ep k \ha   O_t\ep k) \\ \notag &=\sum_i \frac{|O_{ii}|^2}{\mc N^2} + \sum_{i=1}^{N-k} \frac{|O_{i+k, i}|^2}{\mc N^2} \cos[(E_{k+i}- E_i)t)].
\end{align}

If at this point we further assume that we unfold the spectrum, so that the $k$nLS $E_{i+k}-E_i$ does not depend on the density of states $\rho(E_i)$, and average over a suitable ensemble, we find that the time evolution will depend on time only through $f_t\ep k$, which in turn completely determines the $k$nSFF.

The initial operator $\ha O\ep k$ might look somewhat artificial, so let us propose a way to engineer it through a dissipative evolution. In the  case of dissipative dynamics in which the unitary part is dictated by $\ha H_0$ and the dissipator consists of a single Hermitian jump operator $\ha L = \ha L^\dagger$, any system operator evolves according to the \textit{adjoint Lindblad equation} \cite{breuer_theory_2002, rivas_open_2012}
\begin{equation}
    \partial_t \ha O_t = \mr i [\ha H_0, \ha O_t] - \gamma \big[\ha L,[\ha L, \ha O_t]\big],
\end{equation}
where $\gamma$ is the dissipation rate associated with  $\hat L$. Considering commuting operators, $[\ha H_0, \ha L]=0$, which then share a common eigenbasis, $\ha H = \sum_i E_i \ket{i}\bra{i}, \; \ha L = \sum_i l_i \ket{i}\bra{i}$, the solution of the above equation simply reads
\begin{equation}
    \ha O_t = \sum_{i,j} O_{ij} e^{- \mr i (E_i - E_j)t - \gamma (l_i - l_j)^2t} \ket{i}\bra{j}.
\end{equation}
We now assume that we do not apply the Hamiltonian dynamics yet (e.g. going to a rotating frame such that they are not relevant) and that we can engineer the jump operator in a way such that its eigenvalues repeat once after the $k$-th element, namely 
\begin{equation}\label{eq:dissipator}
    \ha L = \mr{diag}(l_1, \dots, l_k, l_1, \dots, l_k, l_{2k + 2},  \dots, l_N), 
\end{equation}
where we have set $l_{k+i}=l_i$ for $1 \leq i \leq k$, and we also consider no extra degeneracies $l_i \neq l_j$ $\forall \; i, j \in \{1, \dots, k, 2k+2, \dots N\}$. The evolution at a time $T$ becomes  $\ha O_T = \sum_{i,j} O_{ij}e^{- \gamma (l_i - l_j)^2 T} \ket{i}\bra{j}$. 
All off-diagonal elements decay exponentially fast with time, except for those with $|i-j|=k$ and $i,j \in \{1, \dots, k\}$ which are preserved due to the structure of $\ha L$. Thus we see that $\lim_{T \rightarrow \infty} \ha O_T = \ha O\ep k = \sum_{i=1}^N \ha O_{ii} \ket{i}\bra{i} + \sum_{i=1}^{2k}O_{i+k, i} \ket{i+k}\bra{i} +\mr{h.c.}$, i.e. this protocol leads to a matrix with $2k$ nonzero elements only in the $k$-th diagonal. The full diagonal could be obtained by repeating the sequence of eigenvalues $l_1, \dots, l_k $ more times in \eqref{eq:dissipator}, but this would lead to higher ``harmonics", i.e. nonzero terms for $|i-j|=2k, 3k, \dots$, which would contain contributions from higher degree $ k$nSFFs.

Other possible experimental probes are to experimentally measure the energy levels of the system, and compute the $k$nLS distribution and the associated $k$nSFF by a Fourier transform. Alternatively, another way could be to use the formalism introduced in \cite{joshi_probing_2022}. More specifically, one would need to find a partition the total Hilbert space in two, $\mathscr H = \mathscr H_A \otimes \mathscr H_B$, such that the condition $\mr{Tr}_B(\rho_B(E_i)\rho_B(E_j))\propto \delta_{|i-j|,k}$ holds, where $\rho_B(E_i) = \mr{Tr}_A(\ket{E_i}\bra{E_i})$. If there exists such a subspace $\mathscr H_A$, then the randomized measurement protocol devised in \cite{joshi_probing_2022} could be readily used to compute the $k$nSFF's and $k$nLS distribution.



%

\end{document}